\documentclass{aa}  

\usepackage{graphicx}
\usepackage{txfonts}
\usepackage{xcolor}
\usepackage{hyperref} 
\usepackage{lastpage}
\begin{document}

   \title{Observing spatial and temporal variations in the atmospheric chemistry of rocky exoplanets: prospects for mid-infrared spectroscopy}
   \titlerunning{Observing 4D atmospheric chemistry of rocky exoplanets with mid-infrared spectroscopy}
   \author{M. Braam
          \inst{1}
          \and
          D. Angerhausen\inst{2,3,4,5}
          % \and 
          % The LIFE Initiative\thanks{\href{https://www.life-space-mission.com/}{https://www.life-space-mission.com/}}
           }

   \institute{Center for Space and Habitability, University of Bern, Gesellschaftsstrasse 6, 3012 Bern, Switzerland\\
              \email{marrick.braam@unibe.ch}
          \and
          SETI Institute, 189 N. Bernado Ave, Mountain View, CA 94043, USA
          \and
          Blue Marble Space Institute of Science, Seattle, 600 1st Avenue, WA 98104, USA
          \and
             ETH Zurich, Institute for Particle Physics \& Astrophysics, Wolfgang-Pauli-Str. 27, 8093 Zurich, Switzerland
             \and
          National Center of Competence in Research PlanetS, Gesellschaftsstrasse 6, 3012 Bern, Switzerland\\             
}

   \date{Submitted to Astronomy \& Astrophysics}

  \abstract
   {Future telescopes such as the Large Interferometer For Exoplanets (LIFE) will enable unprecedented characterisation of the atmospheres of nearby rocky exoplanets, probing mid-infrared signatures of key molecules like CO$_2$, H$_2$O, O$_3$, and CH$_4$. Whilst 4D spatial and temporal variations of Earth as an exoplanet are below spectroscopic detection limits, such variability is strongly planet-specific.}
   {We investigate LIFE's ability to detect 4D spatial and temporal variability in the atmospheres of tidally locked exoplanets.}
   {We create daily synthetic LIFE observations of Proxima Centauri b in a 1:1 and an eccentric 3:2 spin-orbit resonance (SOR), using LIFE\textsc{sim} on spectra from daily 3D climate-chemistry model output of an aquaplanet with Earth-like composition. The spectra assume an inclination of 70$^\circ$.}
   {Hemispheric distributions of temperature, clouds, and chemical species determine spectral signatures and variability with orbital phase angle. Such variability dictates the extent to which parameters (e.g., radius, temperature, or chemical abundances) can be reliably inferred from snapshot spectra at arbitrary viewing geometries. In the 1:1 SOR, MIR spectra vary significantly with viewing geometry and indirectly probe atmospheric circulation. Nightside temperature inversions generate O$_3$, CO$_2$, and H$_2$O emission features, though these lie below LIFE's detection threshold, and instead O$_3$ features disappear at certain phase angles. In contrast, the 3:2 SOR yields a more homogeneous atmosphere with weaker phase variability but enhanced bolometric flux due to eccentric heating. Phase-resolved LIFE observations confidently distinguish between the SORs and capture seasonal O$_3$ variability for golden targets such as Proxima Centauri b. In case of abiotic O$_2$/O$_3$ build-up, the O$_3$ variability presents a potential false positive scenario.}
   {Hence, LIFE can disentangle different spin-orbit states and resolve 4D atmospheric variability, enabling daily characterisation of the 4D physical and chemical state of nearby terrestrial worlds. Importantly, this characterisation requires phase-resolved rather than snapshot spectra.}


   \maketitle
%
%-------------------------------------------------------------------

\section{Introduction}
% General opening paragraph on terrestrial exoplanets and LIFE mission
Whilst the James Webb Space Telescope (JWST) is starting to characterise the atmospheres of favourable rocky exoplanets, the next-generation observatories are being developed and designed to further advance the quest for atmospheric biosignatures. Amongst them are the Extremely Large Telescope (ELT), the Habitable Worlds Observatory (HWO), and the Large Interferometer For Exoplanets (LIFE). LIFE is a space-based nulling interferometer that will observe at mid-infrared (MIR) wavelengths and thereby provide a unique spectroscopic window into the direct thermal emission of rocky exoplanets \citep{LIFE1, glauser2024large}. The thermal emission spectra will probe the atmospheric structure, chemical signatures, and their potential spatial and temporal variations. Therefore, any robust interpretation of the spectra will require a thorough understanding of the stellar and planetary environmental context, especially when distinguishing potential biosignatures from false positive scenarios \citep[e.g.,][]{2018AsBio..18..709C, meadows2018exoplanet}. Here, we evaluate synthetic MIR spectra based on a comprehensive climate-chemistry model to investigate this environmental context and determine LIFE's ability to characterise varying chemical signatures.

Before any such atmospheric characterisation, LIFE is designed to start with a search phase resulting in a detection yield of several hundred exoplanets \citep{LIFE1, LIFE2, LIFE6}. The most promising planets will be targeted for atmospheric characterisation, including planets in M-star habitable zones (typically at $\sim5$~pc) and FGK-star systems (typically at $\sim10$~pc) \citep{LIFE2,LIFE6, LIFE12}. LIFE can characterise an Earth-twin with detected features of ozone (O$_3$), methane (CH$_4$), carbon dioxide (CO$_2$), and water vapour (H$_2$O) \citep{LIFE3} and distinguish between different eras of Earth's geological and atmospheric evolution \citet{LIFE5}. The atmosphere of a Venus-like exoplanet can be characterised above the cloud deck by LIFE, although 1D retrievals struggle to retrieve cloud properties \citep{LIFE9}. More exotic biosignatures, including nitrous oxide (N$_2$O), methylated halogens, and phosphine, are also detectable by LIFE \citep{LIFE8, LIFE12} at plausible biological fluxes and observation times. There are strong synergies between LIFE and HWO, both in terms of planet detection \citep{LIFE10} and robustly characterising atmospheric structure and chemical signatures \citep{life13}, although LIFE has the advantage that it can observe exoplanets on close-in orbits around M stars. Until now, the majority of studies exploring LIFE's characterisation capabilities employed 1D models of atmospheric physics and chemistry.

However, Earth's atmosphere demonstrates substantial variations in its physical and chemical characteristics, in three spatial and one temporal dimension, which in turn lead to 4D variations affecting Earth's (MIR) spectrum \citep[e.g.,][]{des_marais_remote_2002, tinetti_detectability_2006, tinetti_detectability_2006-1, hearty_mid-infrared_2009, gomez-leal_photometric_2012}. Since atmospheric seasonality on Earth is biologically modulated, the temporal (or seasonal) variations have been proposed as a potential biosignature \citep{olson2018atmospheric}. Importantly, observed emission spectra will always be hemispheric averages with unresolved spatial and temporal variations, presenting possible degeneracies in interpretation. \citet{mettler_earth_2023} construct disk-integrated MIR thermal emission spectra of Earth to study the variations of Earth's thermal emission spectrum with viewing geometry, phase angle, and seasons, using remote sensing data of four viewing geometries with high temporal, spatial, and spectral coverage. Moreover, they quantify the atmospheric seasonality of potential biosignatures such as O$_3$, CH$_4$, N$_2$O, and CO$_2$. \citet{mettler_earth_2023} find that `a representative, disk-integrated thermal emission spectrum of Earth does not exist' due to the significant variations with seasons and viewing geometry. The combination of disk-integrated data and annual variability produces degeneracy in the spectrum. To break this degeneracy for Earth, we need more than a single-epoch or snapshot spectrum and cover (part of) the phase angle variations with multiple spectra \citep{mettler_earth_2023}. Atmospheric seasonality in biosignature abundances on Earth generally remains below 5\% (except for CO$_2$ at 8.6 and 15.8\% over the North and South Poles, respectively). 

In a follow-up paper, \citet{mettler_earth_2024} comprehensively assess the potential to detect these biosignatures and their variations as well as the impact of the viewing geometry and seasonality for Earth from 10~pc. They create synthetic 30-day LIFE observations of Earth's MIR spectrum using LIFE\textsc{sim} \citep{LIFE2} and compare subsequent atmospheric retrieval results to the ground truths. LIFE can easily detect a habitable planet (constraining quantities like temperature and albedo) with static detections of CO$_2$, H$_2$O, O$_3$, and CH$_4$, confirming earlier findings \citep{LIFE3, LIFE5}. In case of a well-constrained planet radius, seasonal variations in the surface temperature, equilibrium temperature, and bond albedo are detectable \citep{mettler_earth_2024}. However, the retrieved surface pressure, pressure-temperature profile, and trace abundances have biases, mainly due to assumed constant abundance profiles and the lack of a patchy cloud treatment in the retrieval frameworks, which are both addressed in \citet{konrad2024pursuing}. Earth's spatial and temporal variations in chemical abundances are too small to be detected \citep{mettler_earth_2024}, hindering biosignature detection through seasonality for exoplanets like Earth.

However, many of the currently known rocky exoplanets exhibit orbital configurations that are distinct from Earth's. Since (rocky) exoplanets are most easily detected orbiting close-in to relatively cool stars, the timescales of tidal locking are much shorter than planetary lifetimes, and they end up in spin-orbit resonances or SORs \citep{goldreich_spin-orbit_1966, barnes_tidal_2017, pierrehumbert_atmospheric_2019}. Tidal locking timescales are particularly short for exoplanets around M and K stars and remain shorter than 1~Gyr for K6 stars \citep{barnes_tidal_2017}. The outcomes of tidal locking include various ratios of SOR \citep{goldreich_spin-orbit_1966, dobrovolskis_spin_2007, renaud_tidal_2021}, depending on eccentricity (amongst others). Furthermore, close-in rocky exoplanets may have passed through dynamical cascades of SORs during their evolutionary history \citep{renaud_tidal_2021}, including higher-order resonances. However, the effects for the climatic and dynamical state of a planetary atmosphere are most extreme for the 1:1 and 3:2 SOR, whilst higher-order resonances are more subtle variations of the latter \citep[e.g.,][]{dobrovolskis_spin_2007, colose_effects_2021}. Proxima Centauri b \citep{anglada-escude_terrestrial_2016} is the nearest exoplanet and, located in the Habitable Zone (HZ) of its M5.5V host star, most likely orbits in such a SOR. Furthermore, it is non-transiting, making it ideally suitable for MIR thermal emission spectroscopy. The small distance to Proxima Centauri b (1.032 pc) implies that LIFE can obtain high-quality spectra in observation times on the order of days, making it a `golden target' for the mission with potential characterisation of spatial and temporal variations in its atmosphere \citep{LIFE12}. 
% Paragraph on exoplanet orbital configurations: not like Earth, likely some kind of resonance. Specifically introduce Proxima b. Spatial and temporal variations will be very different. Barnes et al. 2017 paper on timescales and diff star types

A substantial body of theoretical work has hypothesised on the extent of these spatial and temporal variations for Proxima Centauri b and similar exoplanets in SORs, using general circulation models. In the case of a 1:1 SOR, substantial dayside-nightside or longitudinal variations are commonly predicted in meteorological quantities such as irradiation, temperature, and cloud cover \citep[e.g.,][]{turbet_habitability_2016, boutle_exploring_2017, del_genio_habitable_2019, sergeev_atmospheric_2020}. On the other hand, an eccentric 3:2 SOR exhibits a 180$^{\circ}$ shift in the substellar point for every orbit around the star, providing a day-night cycle and latitudinal variations in meteorological quantities \citep{turbet_habitability_2016, boutle_exploring_2017, del_genio_habitable_2019, braam_earth-like_2025}. The 3:2 SOR also generally warms the planet, associated with a weaker cloud feedback and eccentric orbit \citep{colose_effects_2021}. The interplay between stellar irradiation, atmospheric dynamics, thermodynamics, and (photo)chemistry will determine the spatial and temporal distribution of chemical species for either SOR. Studies using coupled climate-chemistry models (CCMs) consistently show substantial dayside-nightside spatial asymmetries in chemical abundances for a 1:1 SOR \citep[e.g.][]{chen_biosignature_2018, yates_ozone_2020, braam_stratospheric_2023, cooke2024lethal}, whereas meridional gradients are expected for a 3:2 SOR \citep{braam_earth-like_2025}. Additionally, temporal variations are caused by a planet's orbital evolution \citep[e.g.,][]{way2017effects, chen2023sporadic, braam_earth-like_2025}, internal atmospheric variability \citep{hochman2022greater, cohen2023traveling, luo2023coupled} or external events such as flares \citep[e.g.,][]{chen2021persistence, ridgway_3d_2023, chen_effects_2025}. Many of the spatial and temporal variations are more pronounced compared to those on Earth.

The spatial and temporal variations, in turn, affect spectroscopic observations of the atmospheres in transmission \citep{chen2021persistence, cohen2023traveling}, reflection \citep{cooke_variability_2023}, and emission spectra \citep{braam_earth-like_2025}. Clearly, variations with seasons or viewing angles depend on the star-planet system, with magnitudes of spectral variations in thermal emission spectra for exoplanets in 1:1 SOR exceeding those of a 3:2 SOR or Earth \citep{mettler_earth_2023, braam_earth-like_2025}. Many false positive scenarios have been reported for static biosignatures, such as the abiotic buildup of O$_2$ and O$_3$ \citep[e.g.,][]{hu2012photochemistry, domagal2014abiotic,tian2014high, harman2015abiotic}. Another robust indicator of biological activity is seasonality in molecular abundances of species like O$_3$ \citep{olson2018atmospheric, schwieterman_exoplanet_2018}. However, if an abiotic pathway to their production exists, `abiotic' spatial and temporal variations due to orbital geometry or internal variability present a potential false positive alternative to seasonal variations in biosignatures that need to be ruled out based on environmental context \citep{fujii2018exoplanet, meadows2018exoplanet}. 

% Then outline the goals of our paper plus structure
In this paper, we investigate LIFE's ability to detect 4D spatial and temporal variability in the atmospheres of tidally locked exoplanets, based on comprehensive 4D climate-chemistry simulations. We aim to distinguish between different SORs and to quantify the spectral variations in O$_3$ due to the planetary and orbital context. In Section~\ref{sec:methods}, we describe the creation of time-resolved synthetic LIFE spectra based on the 3D spatial distributions from the CCM simulations. In Section~\ref{sec:results}, we present the observed spatial distributions, the synthetic LIFE observations, circulation-induced spectral variations, and tie it together in an analysis of temporally varying O$_3$ features. We put our results into context and compare them to Earth in Section~\ref{sec:discussion}, before concluding our study in Section~\ref{sec:conclusion}.

%--------------------------------------------------------------------
\section{Methods}\label{sec:methods}
\subsection{4D Atmospheric Chemistry}\label{subsec:ccm_description}
The data underlying this study were produced using a state-of-the-art CCM --  the Met Office Unified Model in its coupled version to the UK Chemistry Aerosols framework (UM-UKCA) -- and based on the simulations that were analysed in \citet{braam_earth-like_2025}. The modelling framework was initially developed to model Earth's atmosphere and the Earth System \citep[see e.g.,][]{walters_met_2019, archibald_description_2020}.  Here, we briefly discuss the adaptation to terrestrial exoplanets, focusing on the key components of the simulations. For extensive detail on the exoplanet adaptation, we refer the interested reader to \citet{mayne_using_2014, boutle_exploring_2017, yates_ozone_2020, braam_lightning-induced_2022, braam_earth-like_2025}. 

We use UM-UKCA to simulate a 1.1~R$_{\oplus}$ aquaplanet with an Earth-like atmosphere in the orbital configuration of Proxima Centauri b \citep{anglada-escude_terrestrial_2016}. The flat and homogeneous surface is a 2.4~m slab ocean layer without heat transport and has a resolution of 2 by 2.5$^{\circ}$ in latitude and longitude \citep{boutle_exploring_2017}. The Earth-like atmosphere provides a 1~bar surface pressure, and extends up to 85~km in altitude (or a pressure of 9$\times10^{-5}$ or 1.3$\times10^{-4}$ bar, see \citealt{braam_earth-like_2025}). We initialise the atmospheric composition with abundances of N$_2$, O$_2$, and CO$_2$ based on the atmosphere of pre-industrial Earth, and water vapour (or H$_2$O (g)) forms thermodynamically from the surface/atmosphere balance. The radiative transfer scheme (the Suite of Community Radiative Transfer codes based on Edwards and Slingo or SOCRATES, see \citealt{edwards_studies_1996}) interactively calculates heating rates and drives the thermal and dynamical evolution of the atmosphere. The dependence of incoming radiation on the orbital configuration is described in Appendix A of \citet{braam_earth-like_2025}. 

The orbital evolution also determines the photolysis rates of chemical species in the atmosphere, which are calculated in UM-UKCA following \citet{telford_implementation_2013} and the adaptation to exoplanets by \citet{yates_ozone_2020, braam_lightning-induced_2022}. These photolysis rates are the ultimate driver of the (photo)chemistry in the model. The chemistry describes the Chapman mechanism of O$_3$ formation, and the hydrogen oxide (HOx) and nitrogen oxide (NOx) catalytic cycles and is a reduced version of the Stratospheric-Tropospheric scheme presented by \citet{archibald_description_2020}. 

Following \citet{boutle_exploring_2017, braam_earth-like_2025} but also \citet{turbet_habitability_2016} and \citet{del_genio_habitable_2019}, we configure Proxima Centauri b in a 1:1 and 3:2 SOR, including an eccentricity of 0.3 for the latter \citep[see e.g.,][]{goldreich_spin-orbit_1966, dobrovolskis_spin_2007}. For the 1:1 SOR, the star-planet separation is fixed at 0.0485~AU, whereas it varies between 0.03395--0.063~AU for the 3:2 SOR. Furthermore, the rotation rate (or spin) of the 3:2 SOR is increased to 9.7517$\times10^{-6}$ (from 6.501$\times10^{-6}$ for the 1:1 SOR) to cover 1.5$\pi$ rad of planetary rotation in one orbital period of 11.186~days. After spinning up the simulations to a steady state (7400~days for the 1:1 SOR and 17000~days for the 3:2 SOR), we use the first orbit for the 1:1 SOR (12 days of daily output) and the first two orbits for the 3:2 SOR (24 days of daily output). Taking two orbits for the 3:2 SOR covers a full daytime-nighttime cycle for all locations on the planet \citep[see][for details]{braam_earth-like_2025}. To determine the orbital phase angle of the planet, we follow the implementation of orbital astronomy in SOCRATES and the UM \citep{edwards_studies_1996, manners_socrates_2021} and calculate the mean and true anomaly as described in \citet{smart_text-book_1944}. The true anomaly $\nu(t)$ represents the angle between the direction of periastron from the barycenter and the position of a planet at a current time $t$ in days. The mean anomaly $M(t)$ is the fictitious angle from periastron for a planet on a circular orbit at the same time $t$, assuming the same semi-major axis as for the true elliptical orbit and is given by:

\begin{equation}
    M(t) = \frac{2\pi(t-t_p)}{P},
\end{equation}

with $t_p$ the time when the planet is at periastron, and $P$ the length of a year or orbital period (both in days). For a Keplerian orbit of eccentricity $e$, $M(t)$ can be used to calculate the true anomaly $\nu(t)$, using the third-order approximation of a series expansion known as the equation of the centre \citep{smart_text-book_1944}:

\begin{equation}
    \nu(t) = M(t) + \left( 2e-\frac{e^3}{4} \right)\sin(M(t)) + \frac{5}{4}e^2\sin(2M(t)) + \frac{13}{12}e^3\sin(3M(t)).
\end{equation}
For a circular orbit of $e{=}0$, $M(t)$ and $\nu(t)$ are equal. Since we defined the longitude of perihelion in the simulations at 102.94$^\circ$ or 1.796601474~rad, the orbital phase angle $\theta(t)$ as a function of time is found by adding the longitude of perihelion to $\nu(t)$:
\begin{equation}
    \theta(t) = \nu(t) + 102.94.
\end{equation}
$\theta(t)$ and $\nu(t)$ for both the 1:1 and 3:2 SOR can be found in Table~\ref{tab:phase_angles_pcb}. 

The hour angle for the 3:2 spin-orbit resonance (relative to hour angle 0 at perihelion) is given by:
\begin{equation}
    HA(t) = \frac{2\pi(t-t_p)}{t_D} = \frac{2\pi(t-t_p)}{2P},
\end{equation}
where $t_D$ represents the length of day, which is equal to two orbital periods for a 3:2 SOR (and infinite for a 1:1 SOR). The equation of time is used to correct for the discrepancy between the mean and apparent solar (stellar) time when simulating an eccentric orbit and is added as a correction to the hour angle. For the simulations and orbital calculations, we use the equation of time following \citet{mueller_equation_1995, manners_socrates_2021}. Correcting for the clockwise rotation, $2\pi-HA$ then gives $\lambda_{SP}(t)$, the substellar longitude as a function of time. For the 3:2 SOR, $\lambda_{SP}(t)$ is given in Table~\ref{tab:phase_angles_pcb}; for the 1:1 SOR, the substellar point is always at 0$^\circ$ latitude and longitude.

\begin{table}
    \centering
    \caption{True anomaly $\nu(t)$ and orbital phase angle $\theta(t)$ for Proxima Centauri b in a 1:1 and 3:2 spin-orbit resonance (SOR).}
    \begin{tabular}{c|c|c|c|c|c}
        \hline
        {Days} & {$\nu(t)$ 1:1} & {$\theta(t)$ 1:1} & {$\nu(t)$ 3:2} & {$\theta(t)$ 3:2} & {$\lambda_{SP}(t)$ 3:2} \\
        \hline
        1  & 65.33  & 168.26 & 158.63 & 261.57 & 122.21 \\
        2  & 97.51  & 200.45 & 177.21 & {280.15} & 93.42 \\
        3  & 129.69 & 232.63 & 195.16 & 298.10 & 64.98 \\
        4  & 161.88 & 264.81 & 215.63 & {318.57} & 35.43 \\
        5  & 194.06 & 297.00 & 238.79 & 341.73 & 8.53 \\
        6  & 226.24 & 329.18 & 269.04 & {11.98 } & 352.55 \\
        7  & 258.42 & 1.36   & 315.43 & 58.37  & 352.11 \\
        8  & 290.61 & 33.55  & 15.14  & {118.08} & 1.48 \\ 
        9  & 322.79 & 65.73  & 69.87  & 172.81 & 7.37 \\ 
        10 & 354.97 & 97.91  & 107.45 & {210.38} & 359.24 \\
        11 & 27.16  & 130.09 & 133.08 & 236.02 & 336.82 \\ 
        12 & 59.34  & 162.28 & 154.87 & {257.80} & 307.72 \\
        13 &        &        & 173.92 & {276.86 }& 278.67 \\
        14 &        &        & 191.68 & 294.62 & 250.36 \\
        15 &        &        & 211.62 & 314.56 & 220.94 \\
        16 &        &        & 234.20 & 337.14 & 192.96 \\
        17 &        &        & 262.42 & {5.35}   & 174.36 \\
        18 &        &        & 305.43 & {48.36}  & 171.13 \\
        19 &        &        & 3.78   & 106.72 & 179.57 \\
        20 &        &        & 60.89  & 163.82 & 187.14 \\
        21 &        &        & 101.68 & {204.62} & 181.98 \\
        22 &        &        & 128.75 & 231.68 & 161.77 \\
        23 &        &        & 151.00 & 253.94 & 133.25 \\
        24 &        &        & 170.58 & 273.52 & 103.96 \\
        \hline
    \end{tabular}    
    \label{tab:phase_angles_pcb}
\tablefoot{
{We consider one orbit for the 1:1 SOR (11.186 days), and two orbits (22.372 days) for the 3:2 SOR. Spin-up times for both simulations are considered (7400 days and 17000 days for 1:1 and 3:2 SOR, respectively). The substellar point is always at 0$^\circ$ latitude and longitude for the 1:1 SOR. For the 3:2 SOR, the substellar longitude $\lambda_{SP}(t)$ changes over time, as specified. All the angles are given in degrees.}
}
\end{table}

\subsection{PSG spectra}\label{sec:PSG_description}
The Planetary Spectrum Generator (PSG), developed by NASA, is a versatile radiative transfer model suite capable of synthesising planetary spectra across a wide range of wavelengths \citep{villanueva_planetary_2018, villanueva_fundamentals_2022}. For our study, we utilise the Global Emission Spectra (GlobES) application within PSG to generate emission spectra of exoplanets based on the output of 3D climate and chemistry simulations \citep{fauchez2025global}. GlobES has been used to produce spectra of TRAPPIST-1 e as part of the THAI papers \citep{fauchez_trappist-1_2022}, Earth as an exoplanet \citep{kofman_pale_2024}, and Proxima Centauri b \citep{braam_earth-like_2025}.

GlobES creates emission spectra based on 3D distributions of gaseous molecules, including H$_2$O, O$_3$, NO, and NO$_2$, as well as ice and water clouds. We also incorporate iso-abundances of N$_2$, CO$_2$, and O$_2$, along with collision-induced absorption caused by O$_2$-O$_2$, O$_2$-N$_2$, N$_2$-N$_2$, and H$_2$O-H$_2$O pairs. The multiple scattering effects of aerosols, specifically ice and water clouds, and Rayleigh scattering are modelled using the PSGDORT module \citep{villanueva_fundamentals_2022, kofman_pale_2024} and wavelength-dependent extinction coefficients and scattering albedo are calculated using Mie theory. Surface reflection is Lambertian. Full radiative transfer calculations are performed for these 3D distributions taking emission angles into account. A detailed description of the GlobES tool can be found in \citet{kofman_pale_2024, fauchez2025global}.

Our orbital geometry configuration incorporates the standard stellar and orbital parameters for the Proxima Centauri system, assuming an inclination of 70$^\circ$ following \citet{braam_earth-like_2025}.  For the planet in 1:1 SOR, the substellar point is fixed at 0 degrees longitude, with the orbital phase angle varying as shown in Table~\ref{tab:phase_angles_pcb}. For the 3:2 SOR, the substellar longitude and orbital phase angle change over time, following the values in Table~\ref{tab:phase_angles_pcb}. PSG accounts for the planet's eccentricity, resulting in a variable star-planet separation ranging from 0.03395~AU at periastron to 0.06305~AU at apoastron. Standardised configuration files for both SORs and code to convert the CCM data into binary format for the PSG configuration files are available online\footnote{\href{https://github.com/marrickb/LIFE\_4DChem}{https://github.com/marrickb/LIFE\_4DChem}}. For each combination of orbital phase angle, substellar longitude, and climate-chemistry state, PSG calculates the spectral radiance at the top of the atmosphere in units of W~sr$^{-1}$~m$^{-2}$~$\mu$m$^{-1}$. 

\subsection{LIFE\textsc{sim}}
% The Large Interferometer For Exoplanets \cite[LIFE,][]{KammererQuanz2018,LIFE1} is a mission concept for a large, space-based, formation-flying, mid-infrared (MIR) nulling interferometer observatory designed for direct exoplanet detection and atmospheric characterization of terrestrial planets in habitable zones. ESA's Voyage 2050 Senior Committee report (\url{https://www.cosmos.esa.int/web/voyage-2050}) prioritizes ``the characterisation of temperate exoplanet atmospheres in the mid-infrared'' as a prime candidate for future L-class missions. LIFE specifically targets this objective, focusing on planets in M-star habitable zones (typically at $\sim5$~pc) and FGK-star systems (typically at $\sim10$~pc) \citep{LIFE2,LIFE6, LIFE12}.

Following the generation of emission spectra as described in Sections~\ref{subsec:ccm_description} and \ref{sec:PSG_description}, we employ the LIFE\textsc{sim} software to calculate (feature) detectability of the emission spectra, following methodologies from \citet{LIFE8, LIFE12}. Key simulation parameters are summarised in Table~\ref{tab:lifesim}. The reference architecture enables atmospheric characterisation through mid-infrared spectroscopy while balancing sensitivity and technical feasibility \citep{LIFE2,LIFE6,LIFE8}. We assume 24~hours of observation time, matching the temporal resolution of the CCM output for and following earlier assessments of LIFE observations for Proxima Centauri b \citep{LIFE12}.

\begin{table}
\caption{Overview of simulation parameters used in LIFE\textsc{sim}, following \cite{LIFE1} or \cite{LIFE8}.}          
\centering                                     
\begin{tabular}{l l}         
\hline\hline                      
Parameter & Value \\    
\hline                                
    Quantum efficiency & 0.7\\     
    Throughput & 0.05\\
    Minimum Wavelength & 4 $\mu$m       \\
    Maximum Wavelength & 18.5 $\mu$m   \\
    Spectral Resolution & 50      \\
    Interferometric Baseline & 10-100 m \\
    Apertures Diameter & 2 m \\
    Exozodi & 3x local zodi \\
\hline\hline
\end{tabular}\label{tab:lifesim}
\end{table}

\section{Results}\label{sec:results}
We start this section by discussing the observed spatial distributions for different phase angles and their effects on the spectral radiance. We then present simulated LIFE observations and the implications for the observing strategy. The final two subsections outline the prospects for circulation-induced spectral features and seasonally varying biosignatures.

\subsection{Observed spatial distributions}\label{subsec:3d_results}

\begin{figure*}[!ht]
\centering
\includegraphics[width=0.65\linewidth]{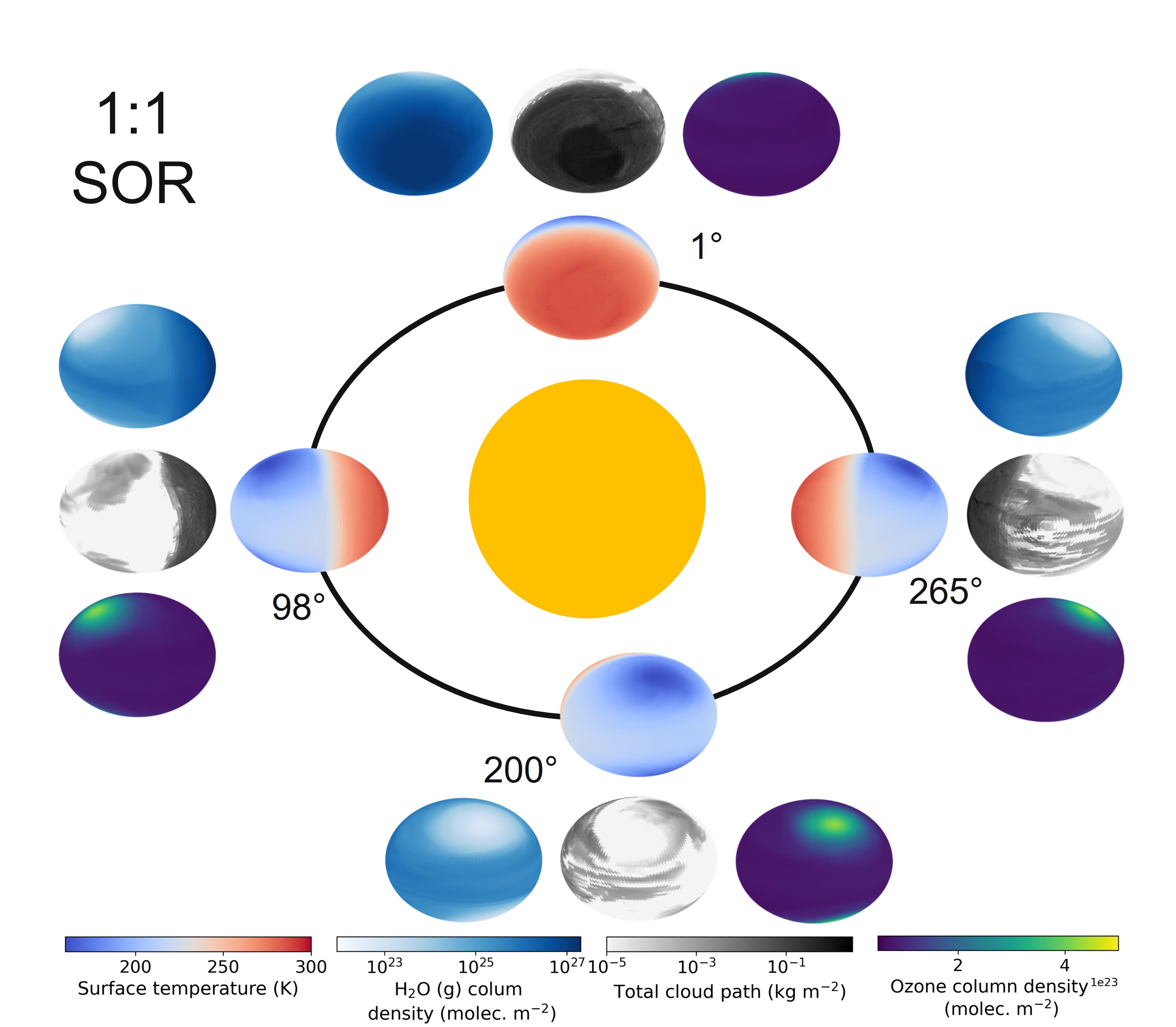}
% \caption{Observed hemispheric distributions as a function of the $\theta(t)$ for the 1:1 SOR: surface temperature (blue-red), vertically integrated H$_2$O (g) column density (white-blue), vertically integrated total cloud path (white-grey), and vertically integrated O$_3$ column density (viridis). The distributions vary both spatially and temporally, as the planet completes one orbit (not to scale). We use four extreme cases of $\theta(t)$ in Table~\ref{tab:phase_angles_pcb}, to investigate the potential impact on observed spectra.}
% \label{fig:pcb_11_3dchem_distrib_phase}
% \end{figure*}
% \begin{figure*}
% \centering
\includegraphics[width=0.63\linewidth]{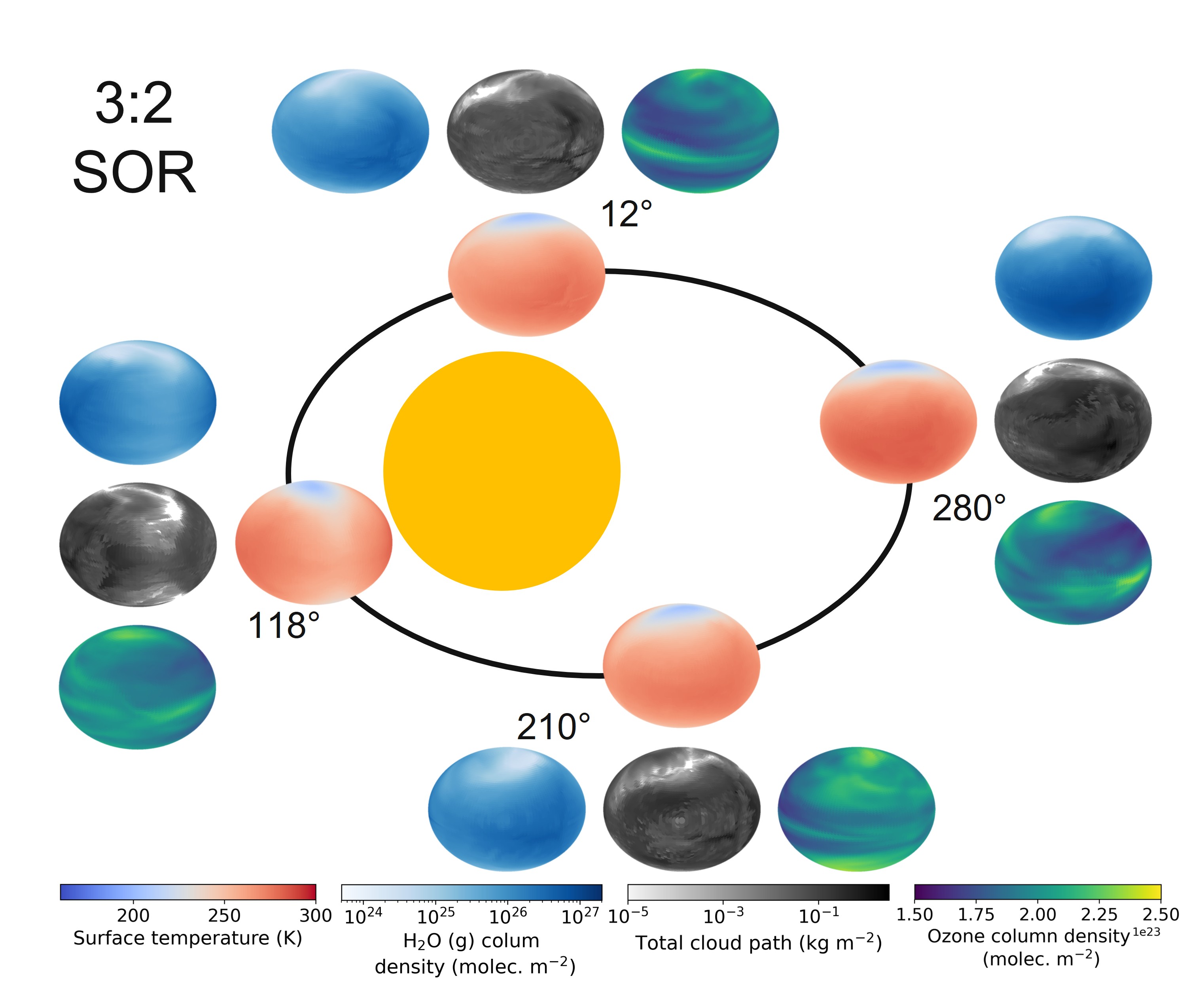}
\caption{Observed hemispheric distributions as a function of $\theta(t)$ for the 1:1 SOR (top) and the 3:2 SOR with an eccentricity of 0.3 (bottom): surface temperature (blue-red), vertically integrated water vapour or H$_2$O (g) column density (white-blue), vertically integrated total cloud path (white-grey), and vertically integrated O$_3$ column density (viridis). The longitude of perihelion of the 3:2 SOR is at 102.94$^\circ$. For both SORs, we use four extreme cases of $\theta(t)$ in Table~\ref{tab:phase_angles_pcb}. The distributions vary spatially and temporally, showing the orbital evolution of the climate and chemistry and effects of viewing geometry.}
\label{fig:pcb_3dchem_distrib_phase}
\end{figure*}

Figure~\ref{fig:pcb_3dchem_distrib_phase} shows how the observed hemispheric distributions of four key quantities change with $\theta(t)$, based on our simulations of Proxima Centauri b in a 1:1 and 3:2 SOR. Shown are: surface temperature $T_{s}$ in K, the vertically integrated water vapour column density $\sigma_{H2O}$ in molec~m$^{-2}$, the vertically integrated total cloud water path (CWP) in kg~m$^{-2}$, and the vertically integrated O$_3$ column density $\sigma_{O3}$ in molec~m$^{-2}$. We include the distributions for four distinct phase angles to demonstrate the extrema in these quantities from an observational perspective. This work aims to connect the predicted 4D spatial and temporal distributions of the Proxima Centauri b simulations to potential observability with LIFE. Therefore, here we only discuss the key characteristics of the observed hemispheric distributions. For more comprehensive descriptions of the physical and (photo)chemical processes underlying the hemispheric distribution, we refer the interested reader to \citet{boutle_exploring_2017} and \citet{braam_earth-like_2025}. 

\begin{figure}
\centering
\includegraphics[width=0.9\linewidth]{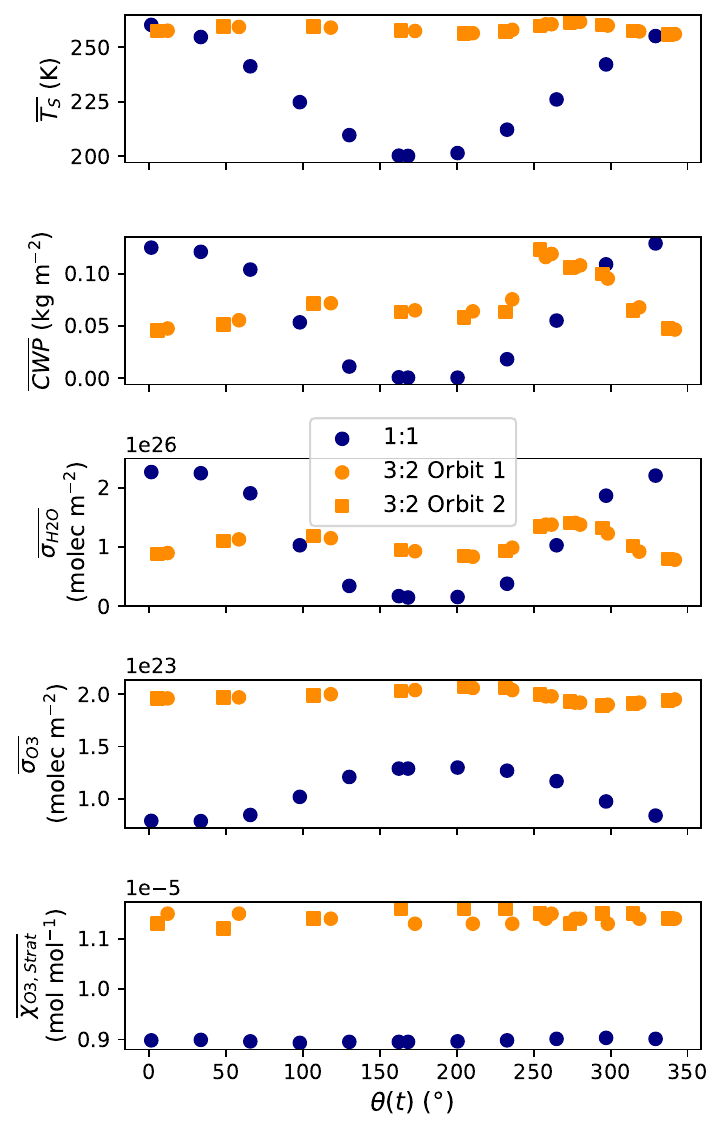}
\caption{Phase angle evolution of the hemispheric means across the observed hemisphere of Proxima Centauri b in 1:1 SOR (navy) and 3:2 SOR (orange), for the quantities shown in Figure~\ref{fig:pcb_3dchem_distrib_phase} and $\overline{\chi_{O3, Strat}}$, the mean volume mixing ratio of O$_3$ in the stratosphere. The hemispheric means are given in Tables~\ref{tab:mean_3d_diagnostics_11} and \ref{tab:mean_3d_diagnostics_32}.}
\label{fig:phase_ccm_diags}
\end{figure}

For each $\theta(t)$, we calculate the hemispheric mean of the four quantities over the hemisphere visible to a distant observer at that time.(see Figure~\ref{fig:phase_ccm_diags}, or Tables~\ref{tab:mean_3d_diagnostics_11} and \ref{tab:mean_3d_diagnostics_32}). Additionally, we calculate the hemispheric mean volume mixing ratio of O$_3$ ($\chi_{O3,strat}$) in the stratosphere, represented by the vertical mean over layers between 100--0.1~hPa.  For the 1:1 SOR, the effects of the synchronous orbit are clearly visible. At $\theta{=}1^\circ$, the observed hemisphere mainly covers the dayside (top panel of Figure~\ref{fig:pcb_3dchem_distrib_phase}), with maxima in $\overline{T_S}$=260.28~K, $\overline{\sigma_{H2O}}{=}$2.27$\times10^{26}$~molec~m$^{-2}$, $\overline{CWP}${=}1.25$\times10^{-1}$~kg~m$^{-2}$, and minima in $\overline{\sigma_{O3}}{=}$ 7.91$\times10^{22}$~molec~m$^{-2}$ (Figure~\ref{fig:phase_ccm_diags}). Close to first quadrature, at $\theta{=}98^\circ$, Figure~\ref{fig:pcb_3dchem_distrib_phase} shows that the observed hemisphere is comprised of both the dayside and nightside hemispheres, with lower $\overline{T_S}$=224.82~K, $\overline{\sigma_{H2O}}${=}1.03$\times10^{26}$~molec~m$^{-2}$, and $\overline{CWP}${=}5.35$\times10^{-2}$~kg~m$^{-2}$, as well as higher $\overline{\sigma_{O3}}{=}$ 1.02$\times10^{23}$~molec~m$^{-2}$. This trend is continued until conjunction, where we mainly observe the nightside hemisphere as illustrated in Figure~\ref{fig:pcb_3dchem_distrib_phase} for $\theta{=}200^\circ$. Now, the averages on the observed hemisphere reach minima in $\overline{T_S}$=201.46~K, $\overline{\sigma_{H2O}}${=}1.54$\times10^{25}$~molec~m$^{-2}$, and $\overline{CWP}${=} 5.98$\times10^{-4}$~kg~m$^{-2}$ and maxima in $\overline{\sigma_{O3}}{=}$1.30$\times10^{23}$~molec~m$^{-2}$. Lastly, the third quadrature shows part of the dayside and part of the nightside hemisphere again, similarly to the first quadrature and reflected in the observed hemispheric averages in Figure~\ref{fig:phase_ccm_diags} and Table~\ref{tab:mean_3d_diagnostics_11}. Figure~\ref{fig:phase_ccm_diags} shows how $\chi_{O3,strat}$ only varies by up to 1\%. These small variations agree with a long chemical lifetime for stratospheric O$_3$ (10--1000 years) as compared to an orbital period \citep{braam_stratospheric_2023}.
%Since the total O$_3$ column density is mostly sensitive to the troposphere -- the atmospheric layer with the highest air density -- the trends of $\overline{\sigma_{O3}}$ with $\theta(t)$ might not be reflected in the emission spectra.

For the 3:2 SOR, the bottom panel of Figure~\ref{fig:pcb_3dchem_distrib_phase} illustrates the effect of a non-synchronous tidally locked orbit. Due to the enhanced planetary spin velocity and thus the absence of a permanent dayside hemisphere, the distributions of $\overline{T_S}$, $\overline{\sigma_{H2O}}$, $\overline{CWP}$, and $\overline{\sigma_{O3}}$ are more homogeneous (compared to the 1:1 SOR). This enhanced homogeneity is reflected in the hemispheric averages for the observed hemispheres in Figure~\ref{fig:phase_ccm_diags}. We find maxima in $\overline{T_S}$=261.74~K, $\overline{\sigma_{H2O}}${=}1.41$\times10^{26}$~molec~m$^{-2}$, and $\overline{CWP}${=}1.06$\times10^{-1}$~kg~m$^{-2}$ and minima in $\overline{\sigma_{O3}}{=}1.89\times10^{23}$~molec~m$^{-2}$ for $\theta{\sim}280^\circ$. In this case, we observe the side of the planet that was previously subjected to the strongest radiation during the daytime at periastron passage. On the other hand, we find minima in $\overline{T_S}$=255.90~K, $\overline{\sigma_{H2O}}${=}7.85$\times10^{25}$~molec~m$^{-2}$, and $\overline{CWP}${=}4.58$\times10^{-2}$~kg~m$^{-2}$ for $\theta{\sim}0^\circ$ (observing the side that was illuminated at apoastron) and maxima in $\overline{\sigma_{O3}}{=}1.89\times10^{23}$~molec~m$^{-2}$ slightly earlier in the orbit for $\theta{=}295^\circ$. Stratospheric variations are enhanced compared to the 1:1 SOR, with $\chi_{O3,strat}$ variations up to 3.6\%.

\begin{figure*}
\centering
\includegraphics[width=0.48\linewidth]{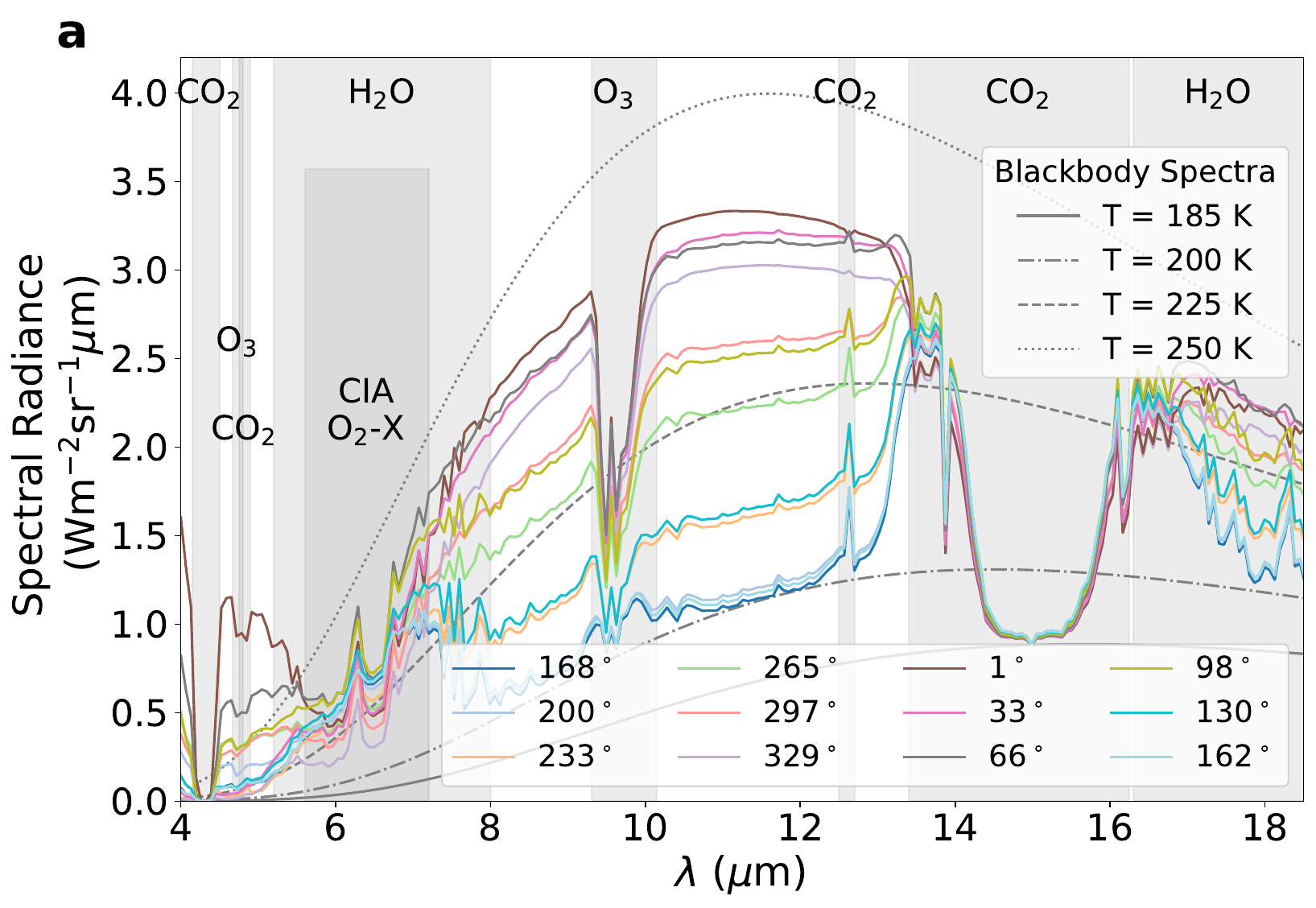}
\includegraphics[width=0.48\linewidth]{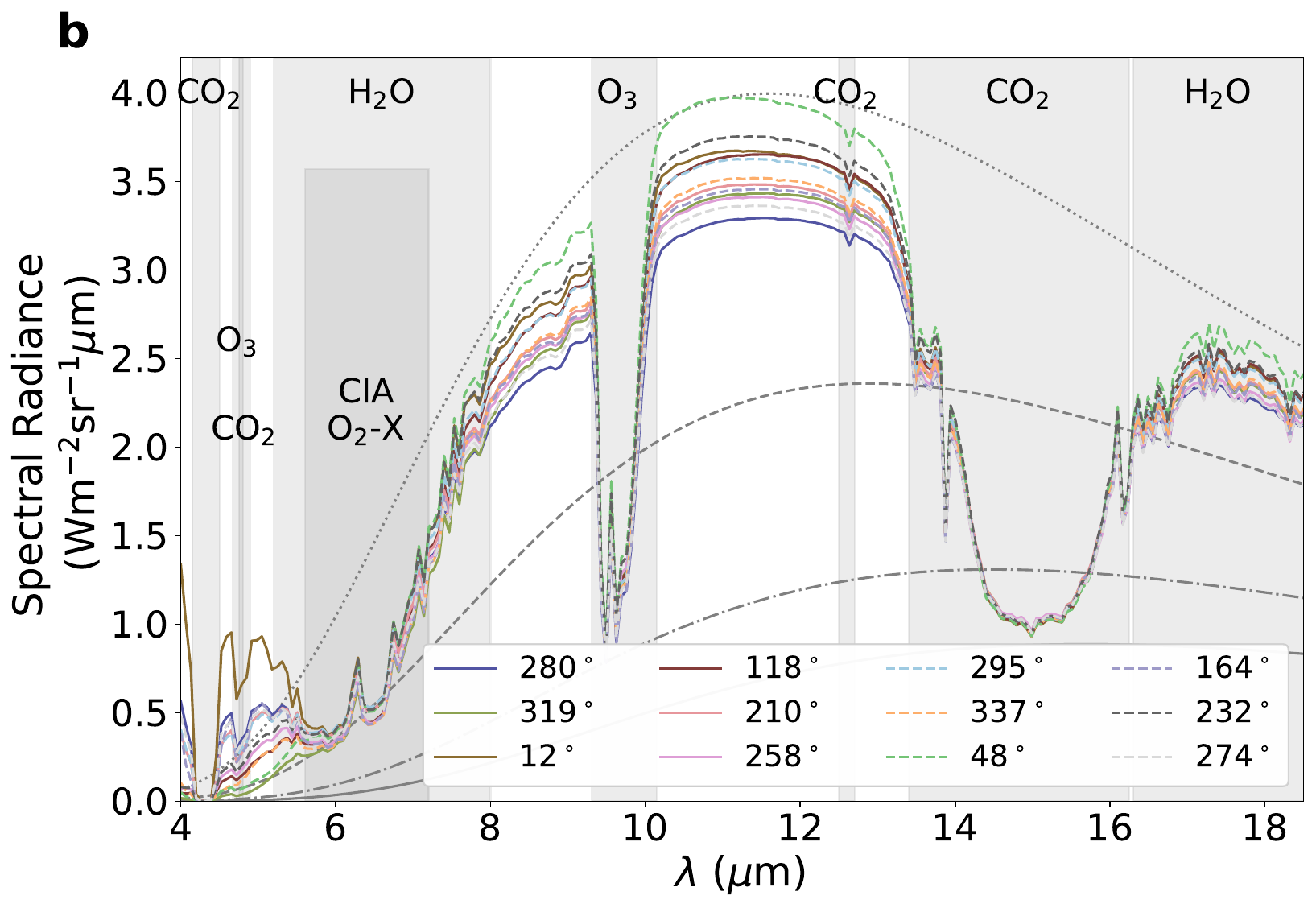}
\caption{Simulated spectral radiance at the top of the atmosphere for Proxima Centauri b in (a) 1:1 SOR and (b) 3:2 SOR, for $\theta(t)$ as shown in Table~\ref{tab:phase_angles_pcb}. Spectra were created using PSG and the GlobES tool (see Section~\ref{sec:PSG_description} for details). We also include blackbody curves at different temperatures for comparison and grey rectangles for important molecular and collision-induced absorption (CIA) features. For panel (b), the solid and dashed lines represent the first and second orbits around the host star, respectively, and together correspond to the length of a full day for the 3:2 SOR.}
\label{fig:pcb_emission_spectra}
\end{figure*}

Figure~\ref{fig:pcb_emission_spectra}a shows the simulated emission spectra for Proxima Centauri b in a 1:1 SOR, as modelled by the GlobES tool of PSG (see Section~\ref{sec:PSG_description}). For the 1:1 SOR, we include daily spectra to show the dependence on $\theta(t)$. The maximum continuum emission (as probed between e.g. 10--12.5~$\mu$m) is received for $\theta{=}1^\circ$, in line with Figure~\ref{fig:pcb_3dchem_distrib_phase} where we observe most of the dayside hemisphere. The planet then goes through phase angles for which we observe a combination of the dayside and nightside hemispheres, leading to lower emission, until the point of minimum emission, when mainly the nightside hemisphere is observed (e.g., at $\theta$=162, 168 or 200$^\circ$). Compared to the blackbody curves, the continuum varies from approximately in the middle of the 225 and 250~K blackbodies down to the 200~K curve. The CO$_2$ feature between 14--16~$\mu$m originates from the relatively cool stratosphere and therefore even approaches the 185~K blackbody curve.

The absorption due to the main O$_3$ feature (around 9.6~$\mu$m) is especially deep when most of the dayside is in view. As we transition to observing the nightside hemisphere O$_3$ absorption is still present but less prominent (e.g., at $\theta$=162, 168, or 200$^\circ$). Additionally, the outer regions become emission features, likely due to the formation of a tropospheric inversion layer \citep[e.g.,][]{2020ApJ...893..140G}. The varying depth and behaviour of the O$_3$ feature suggest that a statistically significant O$_3$ detection might be dependent on the observed phase angle. Similarly, the 4.8~$\mu$m O$_3$ feature is deepest when the dayside is most into view. However, the generally shallower feature at this wavelength completely disappears for all but the brightest phase angles. The features due to collision-induced absorption (CIA) of O$_2$-O$_2$ and O$_2$-N$_2$ pairs (between 5.2--7.4~$\mu$m) are mainly overshadowed by the stronger H$_2$O features in the same region. Nevertheless, the narrow peak in the middle (centred at 6.4~$\mu$m) might be an O$_2$ signature as long as the phase angle dependence of H$_2$O features is known.  

The H$_2$O features (5--8 and 16.5--18.5~$\mu$m) vary as expected: with the hottest and thus wettest observed hemispheres ($\theta$=1, 33, 66, 329$^\circ$) resulting in the strongest emission. Transitions between absorption and emission features are seen at 7.9 (H$_2$O), 12.6 (CO$_2$), and 18.3~$\mu$m (H$_2$O). When we mainly observe the dayside hemisphere, we tend to see absorption as compared to surrounding wavelengths, which changes to emission when we mainly observe the nightside. These features originate in the troposphere and are thus affected by the nightside tropospheric temperature inversion: they originate at higher temperatures than the surrounding features \citep[e.g.,][]{2020ApJ...893..140G}. 

For the 3:2 SOR in Figure~\ref{fig:pcb_emission_spectra}b, the more homogeneous atmosphere substantially reduces the phase angle variations. Furthermore, the higher spin rate and eccentric orbit affect the appearance of maxima and minima in the spectral radiance. As shown in Table~\ref{tab:mean_3d_diagnostics_32}, observed hemispheric temperature maxima are found at phase angles close to apoastron (e.g., 250--290$^\circ$), when the hemisphere that was subjected to periastron irradiation comes into view. However, $\overline{CWP}$ is also twice as high on the observed hemispheres around apoastron compared to low phase angles (e.g., 12 and 48$^\circ$), offsetting the relatively small temperature changes. The maximum continuum emission in Figure~\ref{fig:pcb_emission_spectra}b comes from $\theta{=}48^\circ$, due to this combined effect of planetary temperature and cloud thickness. The continuum emission is closest to the 250~K blackbody curve, regardless of phase angle, due to the warmer atmosphere of a 3:2 SOR compared to a 1:1 SOR. The 9.6~$\mu$m O$_3$ and 14--16~$\mu$m CO$_2$ features originate in the cooler stratosphere, thus approaching the 200~K and 185~K curves, respectively. Phase angle variations in individual chemical signatures (like O$_3$ or H$_2$O) are minimal in this case, demonstrating less potential ambiguity in detecting these species.

%plot with average vertical T profile and range of all vertical profiles with orbital phase and planet location?
%Emission spectra of the extreme cases, overplotting simulated LIFE observations for 1:1 and 3:2 case
%Then perhaps investigate amount of hours needed for detections, 3D effects, distinguishing 1:1 versus 3:2
%or perform a few retrievals based on iso-abundances versus physical profiles (Konrad et al., 2024)

\subsection{LIFE observations \& consequences for observation strategy}\label{subsec:life_spectra}
\begin{figure*}
    \centering
    \includegraphics[width=0.48\linewidth]{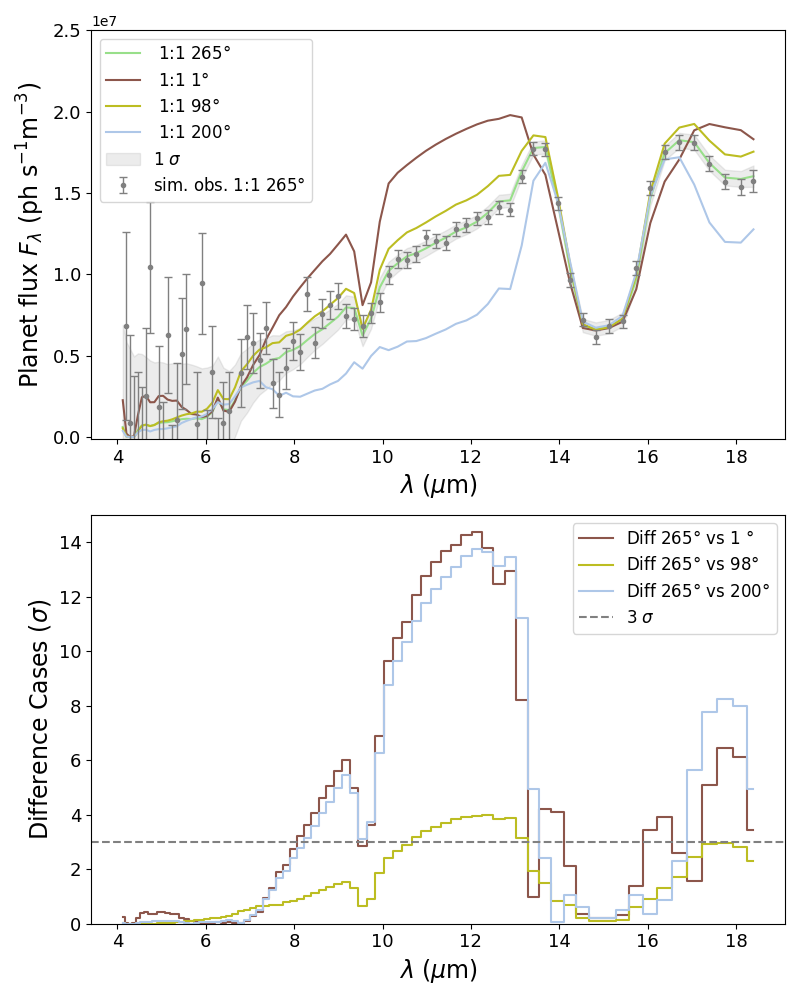}
    \includegraphics[width=0.48\linewidth]{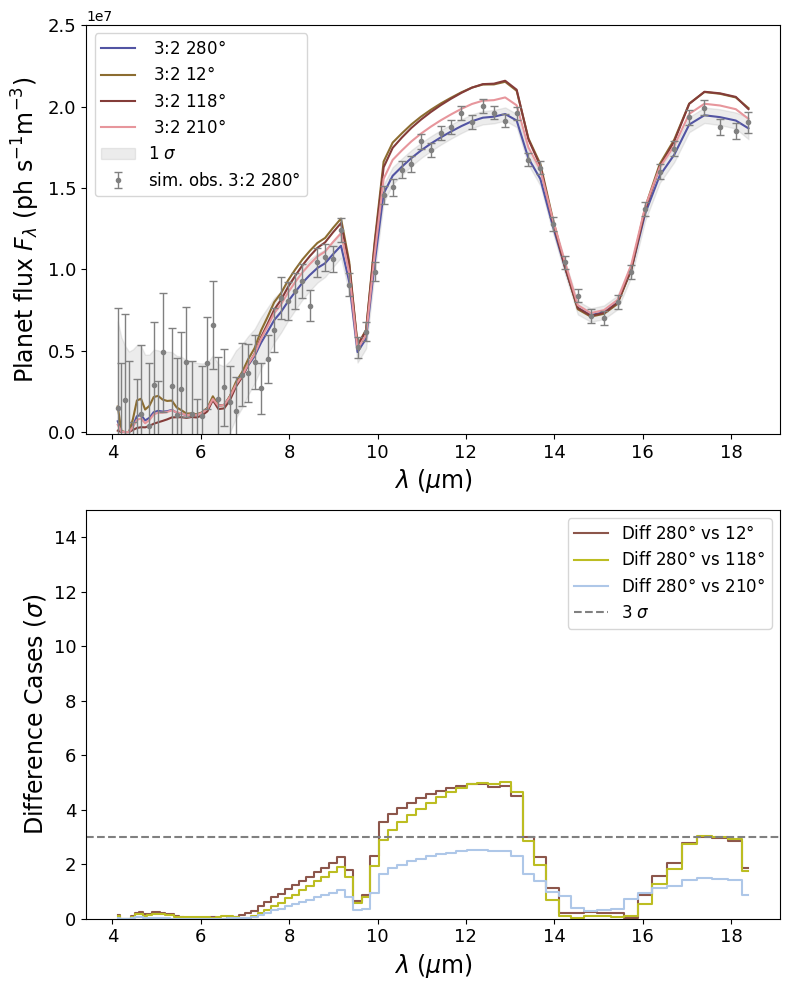}

    \caption{Example simulated LIFE observation for the 1:1 resonance case (left) and 3:2 resonance case (right) assuming an integration time of 24 hours. The grey area represents the 1-$\sigma$ sensitivity; the dark grey error bars show an individual simulated observation. Lower panel: Statistical significance of the detected differences between different phases.}
    \label{fig:LIFE_spectra_all}
\end{figure*}

%If detection observation with LIFE, how much would we know about this planet already? Detection-phase focused, spread in temperatures
Proxima Centauri b has previously been identified as a golden target for LIFE, due to its relative proximity to the Solar System \citep{LIFE12}, potentially allowing for detailed and time-resolved observations of the planet. \citet{LIFE12} test the observability of a 'static' Earth-like exoplanet with varying biogenic fluxes, and find that detailed observations are already possible with just 24~hours of observational time with LIFE. Here, we expand upon their results by employing LIFE\textsc{sim} \citep{LIFE2} to create synthetic time-resolved observations with LIFE based on the CCM simulations in Section~\ref{subsec:3d_results}. We also assume 24~hours of observational time with LIFE, in agreement with the temporal resolution of CCM output. Figure~\ref{fig:LIFE_spectra_all} shows the predicted LIFE observations in a phase comparison for Proxima Centauri b, comparing LIFE observations at distinct phase angles for the 1:1 SOR (left) and the 3:2 SOR (right).

Focusing on the observations for the 1:1 SOR at $\theta{=}265^{\circ}$, we first see that LIFE can conduct a detailed characterisation of the spectrum for a specific phase angle in one Earth day. LIFE clearly captures the O$_3$ and CO$_2$ features and is also sensitive to the extent of the H$_2$O features between 16.5--18.5~$\mu$m. Second, LIFE is highly sensitive to the spectral changes with $\theta$ for the 1:1 SOR. The statistical significances of observed differences between phase angles are shown in the bottom panels of Figure~\ref{fig:LIFE_spectra_all}. For the 1:1 SOR, the statistical differences reach 12~$\sigma$ in the continuum level and ${\sim}8\sigma$ for the H$_2$O features. Moreover, compared to $\theta{=}1^{\circ}$ or $98^{\circ}$, the observed differences around the 9.6~$\mu$m O$_3$ feature are significant near the $3\sigma$ level. Clearly, daily LIFE observations will provide detailed information on changes in the chemical state (O$_3$ and H$_2$O abundances and distributions) and physical state (temperature, clouds, and H$_2$O distributions) of the planetary atmosphere. Hence, the high spectral and temporal resolution with LIFE observations of Proxima Centauri b allow us to probe the dynamic atmosphere on a daily basis.

The comparison with the 3:2 SOR in the top left panels of Figure~\ref{fig:LIFE_spectra_all} confirms the eccentricity-induced global temperature enhancement for the 3:2 resonance, with generally higher continuum fluxes than the 1:1 SOR. Even though the atmosphere of the 3:2 SOR is more homogeneous (see Section~\ref{subsec:3d_results}), LIFE observations still vary with $\theta$ at up to ${\sim}5\sigma$ significance, providing a potential probe to the extremes of the dynamic atmosphere given LIFE's spectral and temporal resolution. Nevertheless, the emission differences with $\theta$ are considerably smaller for the 3:2 SOR, providing a clear distinction from a planet in a 1:1 SOR. The low nightside continuum emission for the 1:1 (see the $\theta{=}168^{\circ}$) is key to making this distinction. The homogeneous distribution of photochemical species such as O$_3$ for a 3:2 SOR makes the spectral variations even smaller at the relevant absorption wavelengths (see the 9.6~$\mu$m features). Therefore, a phase curve centred at the O$_3$ feature (or features of other prominent photochemical species) is an excellent probe of spin-orbit resonances.

% Phase curves as "signal"

% \begin{itemize}
%     \item example for "golden target" $\checkmark$
%     \item LIFE able to do detailed phase analysis of Prox B $\checkmark$
%     \item high spectral-temporal resolution enables observing "dynamic" planets $\checkmark$
%     \item LIFE can clearly distinguish tidal locked from 3:2 resonance case $\checkmark$
%     \item not only for Prox B but a substantial (to be at least ballpark quantified) number of late type systems withing $\sim$ 5-7 parsec
%     \item motivates further, more detailed development of (3D) models as this study shows that effects are observable
%     with LIFE
% \end{itemize}

%What is the distance dependence? 
%can we distinguish between different orbital configurations
%combine phases, what happens to longer integration times?

\subsection{Atmospheric circulation mechanisms}\label{subsec:circulation}
Beyond the orbital configuration, the spatial distributions of temperature, clouds, and chemical species are strongly affected by the atmospheric dynamics on a planet. The 1:1 versus 3:2 SOR comparison in Figure~\ref{fig:pcb_3dchem_distrib_phase} illustrates this most evidently in the O$_3$ column density. Despite being photochemically produced, O$_3$ accumulates on the nightside of the 1:1 SOR and at high latitudes for the 3:2 SOR due to their respective circulation regimes \citep{braam_earth-like_2025}. The spectral variations in simulated LIFE observations for the 3:2 SOR are too small to provide us with clues to the specific circulation regime, especially since chemical abundances vary with planetary latitude. The longitudinal variations for a 1:1 SOR leave more promising phase angle variations in the LIFE observations relating to the circulation regime.

For exoplanets in 1:1 SOR, the specific regime of atmospheric circulation is predicted to depend on the orbital period \citep[e.g.][]{carone_connecting_2015, noda_circulation_2017, haqq-misra_demarcating_2018} as well as model parametrisations \citep{sergeev_bistability_2022}. The different circulation regimes for a 1:1 SOR will leave their marks in spatial distributions of temperature, clouds, water vapour and chemical abundances \citep[e.g.,][]{chen_habitability_2019, sergeev_trappist-1_2022, braam_stratospheric_2023}. Proxima Centauri b simulations commonly exhibit an eastward equatorial jet in the troposphere. In Figure~\ref{fig:pcb_3dchem_distrib_phase}, we see how the circulation regime manifests itself in higher $\overline{CWP}$ for a phase angle of 265$^{\circ}$ as compared to 98$^{\circ}$, as the eastward advection of the substellar cloud deck comes into view. The same mechanism might apply to two other pairs of phase angles: $233^{\circ}$ versus 130$^{\circ}$ and 297$^{\circ}$ versus 66$^{\circ}$, although these contaminations from dayside-nightside differences are bigger due to a larger angular distance from 90$^{\circ}$ and 270$^{\circ}$.

\begin{figure}
\centering
\includegraphics[width=0.85\linewidth]{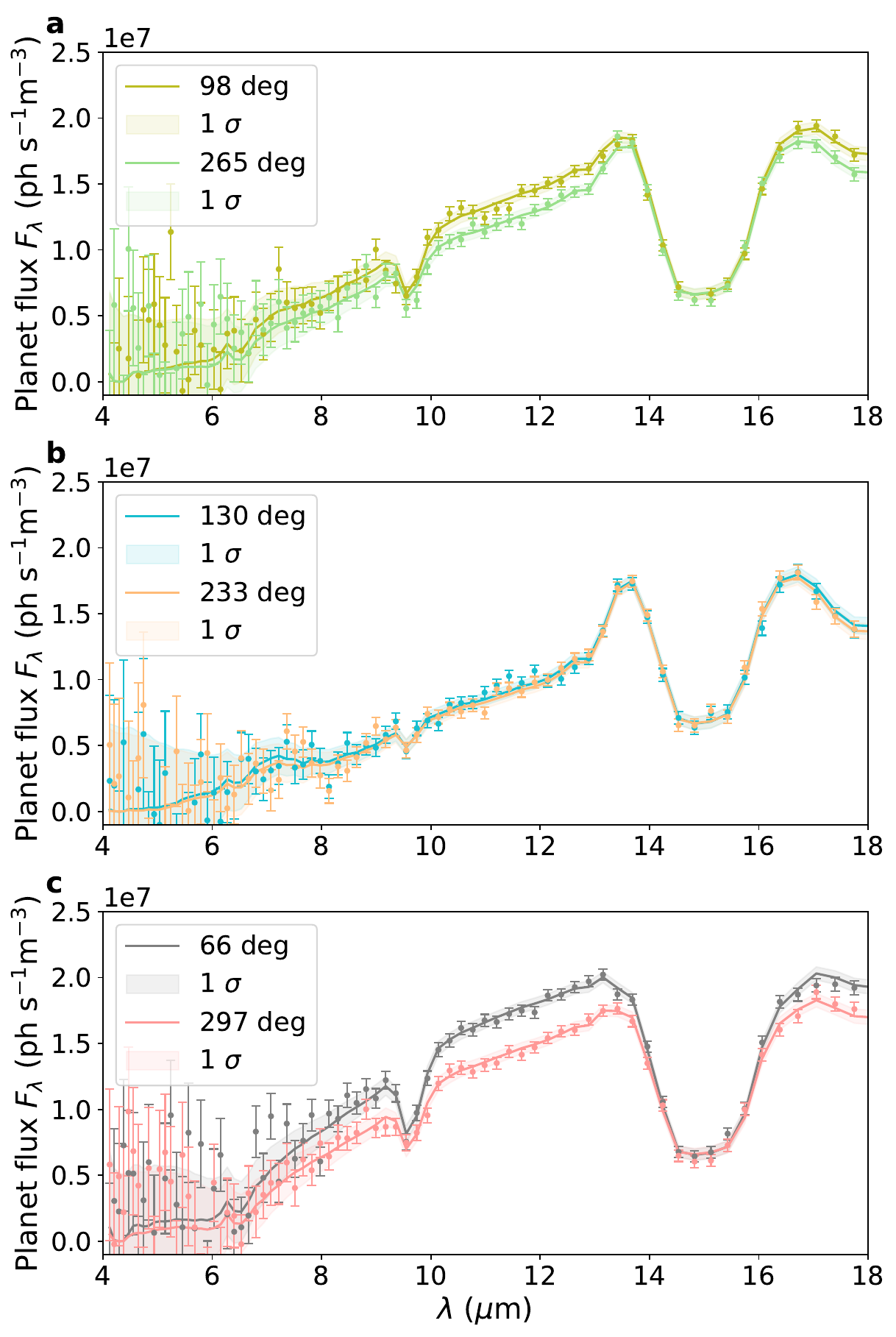}
\caption{Pairs of simulated LIFE spectra for the 1:1 SOR at similar distances from phase angles of 90 and 270$^{\circ}$, illustrating the effects of the atmospheric circulation. Panel a shows spectra for phase angles 98 and 265$^{\circ}$, panel b for 130 and 233$^{\circ}$, and panel c for 66 and 297$^{\circ}$.}
\label{fig:pcb_spectra_circ}
\end{figure}

We compare these phase angle pairs in the three panels of Figure~\ref{fig:pcb_spectra_circ}. The planet fluxes observed by LIFE are enhanced for a phase angle of 98$^{\circ}$ compared to 265$^{\circ}$ (Figure~\ref{fig:pcb_spectra_circ}a). Hence, despite the slightly lower $\overline{T_S}$ (0.6\%, see Table~\ref{tab:mean_3d_diagnostics_11}), the 3\% lower $\overline{CWP}$ allows planet flux to originate from deeper (and warmer) atmospheric layers. The flux differences are markedly weaker when comparing phase angles 130$^{\circ}$ and 233$^{\circ}$ (Figure~\ref{fig:pcb_spectra_circ}b). For these phase angles, LIFE mainly observes the nightside hemisphere with its horizontal asymmetries due to the Rossby gyres and vertical temperature inversions. Moreover, the actual values for $\overline{CWP}$ are five times lower than phase angles 98$^{\circ}$ and 265$^{\circ}$, presenting too little cloud cover to be sensitive to the circulation mechanism. For phase angles 66$^{\circ}$ and 297$^{\circ}$ and Figure~\ref{fig:pcb_spectra_circ}c), LIFE again observes more planet flux for the smaller phase angle, which presents the lower $\overline{T_S}$ (0.3\%) and $\overline{CWP}$ (5\%). Moreover, $\overline{CWP}$ is at similar levels to phase angles 98$^{\circ}$ and 265$^{\circ}$, giving enough cloud cover to provide sensitivity to the circulation mechanism. Hence, deliberate phase-resolved emission spectroscopy with LIFE can be used to confirm the predicted circulation mechanisms through the advection of the substellar cloud deck.

\subsection{Temporally varying biosignatures}\label{subsec:seasonality}
%Seasonal varations in relation to seasonally varying biosignatures
%orbital phase importance in looking for biosignatures
In Sections~\ref{subsec:3d_results} and \ref{subsec:life_spectra} we identified different temporal variations due to the stellar and planetary environment and, in our work, in particular due to distinct SORs. Evidently, temporal variations in emission spectra arise due to 1) seasonality on a planet, driven by obliquity or eccentricity, and 2) orbital phase angle and viewing geometry in the case of spatial asymmetries in the planetary atmosphere. Since seasonality has previously been proposed as a biosignature, a thorough understanding of the robustness of seasonality as a biosignature is essential -- particularly in the context of abiotic O$_2$/O$_3$ buildup on exoplanets around M-dwarfs \citep[e.g.,][]{hu2012photochemistry, domagal2014abiotic,tian2014high, harman2015abiotic}. \citet{mettler_earth_2023} use time-series of remote sensing data of Earth's atmosphere to show how Earth's MIR spectrum varies with time, viewing geometry, and phase angle, focusing on absorption features of O$_3$, CH$_4$, CO$_2$, and N$_2$O. In our simulations, CO$_2$ is fixed and CH$_4$ absent, whereas abiotically produced N$_2$O (lightning, photochemistry) is not abundant enough to be detectable \citep{braam_lightning-induced_2022}. The 9.6~$\mu$m O$_3$ feature, however, shows substantial variations (see Figures~\ref{fig:pcb_emission_spectra} and \ref{fig:LIFE_spectra_all}). Hence, we can investigate the most physically and chemically self-consistent time series of O$_3$ variations, for the two SORs considered, given our model assumptions.

\begin{figure}
\centering
\includegraphics[width=0.72\linewidth]{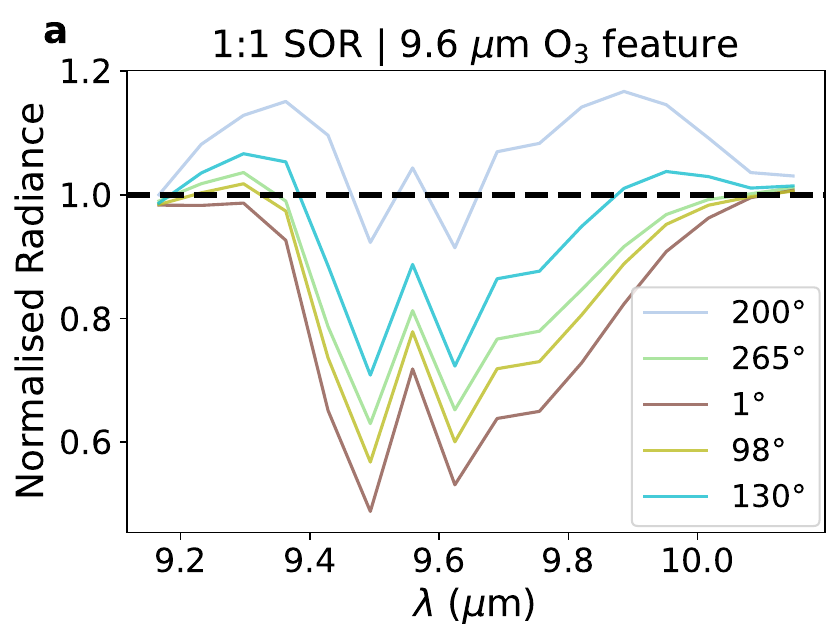}
\includegraphics[width=0.72\linewidth]{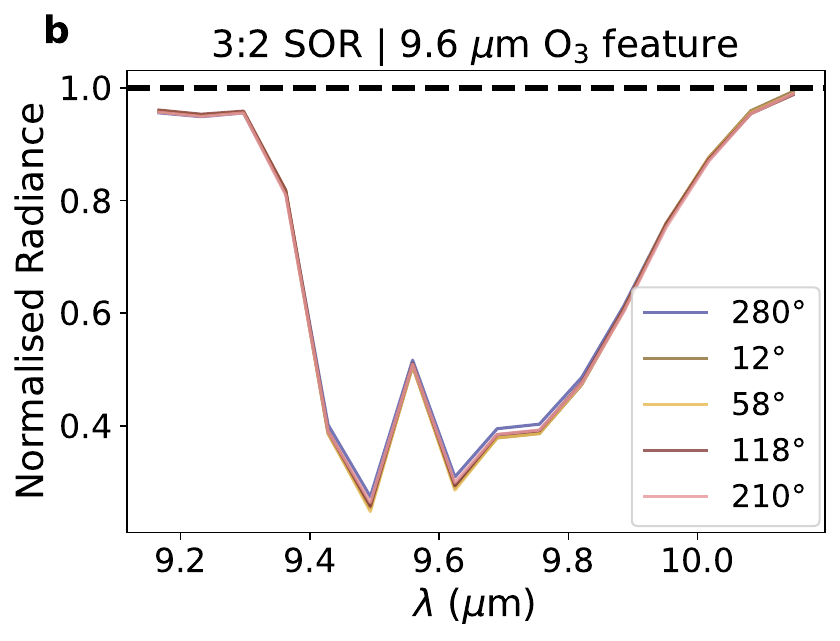}
\caption{Normalised radiance centered on the 9.6~$\mu$m O$_3$ features, for selected phase angles of the 1:1 SOR (a) and 3:2 SOR (b). We fit the continuum data (8--9.15 and 10.15--12.2~$\mu$m) with a quadratic polynomial and normalise the radiance. This approach allows us to isolate the O$_3$ feature as a function of orbital phase angle, illustrating the change from absorption to emission for the 1:1 SOR and the generally stronger absorption features for the 3:2 SOR.}
\label{fig:pcb_o3_normrad}
\end{figure}

We follow the approach by \citet{mettler_earth_2023} and calculate the equivalent width as a measure of the strength of an absorption or emission feature:
\begin{equation}\label{eq:equiv_width}
    W_\lambda = \int \left(1-\frac{I_{O_3}}{I_C}\right)d\lambda,
\end{equation}
where $I_{O_3}$ is the radiance in our region of interest (in this case, the O$_3$ feature at 9.6~$\mu$m, using outer bounds of 9.15--10.15~$\mu$m) and $I_{C}$ the continuum radiance in this same region, based on a fit to the wavelength regions surrounding the 9.6~$\mu$m O$_3$ feature, specifically between 8--9.15~$\mu$m and 10.15--12.2~$\mu$m. We fit the continuum regions with a quadratic polynomial and normalise the radiance in the O$_3$ band with the fitted continuum radiance, as in Equation~\ref{eq:equiv_width}. Figure~\ref{fig:pcb_o3_normrad} demonstrates how this procedure allows us to isolate the O$_3$ features, for both SORs. Note that our O$_3$ bands are wider than the 0.7~$\mu$m band used by \citet{mettler_earth_2023}, motivated by the broad features originating from the nightside phase angles of the 1:1 SOR. Since these features emerge in the troposphere, they are likely subject to pressure broadening. The continuum fit slightly overestimates the lower wavelength continuum level for the 3:2 SOR in Figure~\ref{fig:pcb_o3_normrad}b, giving a normalised radiance below unity. However, it does not affect variations in W$_\lambda$ since the normalised radiance is constant between 9.15--9.3~$\mu$m. The homogeneous atmosphere of the 3:2 SOR produces a fairly constant normalised radiance (Figure~\ref{fig:pcb_o3_normrad}b). On the other hand, the normalised radiance in the O$_3$ band varies considerably for the 1:1 SOR. Notably, the 1:1 SOR shows the transitions between absorption and emission, with a normalised radiance smaller and greater than unity, respectively. The nightside phase angles (168 and 200$^\circ$) again correspond to O$_3$ emission features that are due to the nightside temperature inversion (see Section~\ref{subsec:3d_results}). 

\begin{figure}
\centering
\includegraphics[width=0.82\linewidth]{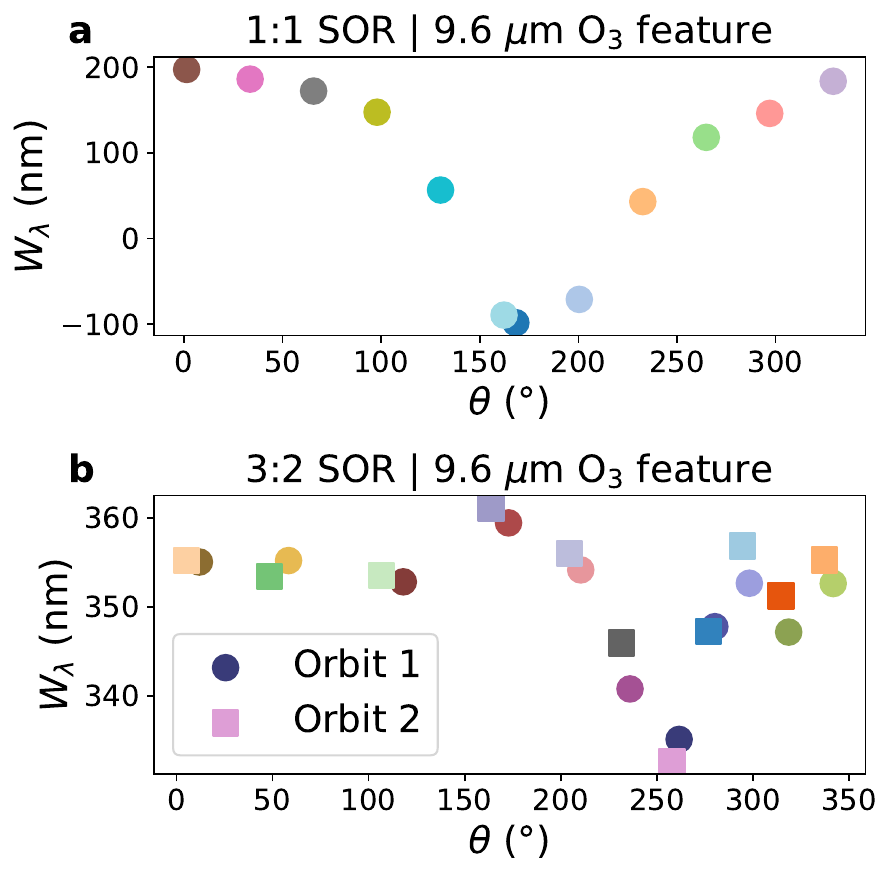}
\caption{Equivalent width (W$_\lambda$) as calculated by Equation~\ref{eq:equiv_width} as a function of the orbital phase angle $\theta$, for the 1:1 SOR (a) and 3:2 SOR (b). The colours correspond to the curves shown in Figure~\ref{fig:pcb_o3_normrad} and we separate the two orbits of the 3:2 SOR with distinct markers. The evolution of W$_\lambda$ with $\theta$ shows how different orbits provide distinct temporal variations in biosignatures, attributed to viewing geometry and seasonality.}
\label{fig:pcb_o3_ew}
\end{figure}

\begin{figure*}
\centering
\includegraphics[width=0.8\linewidth]{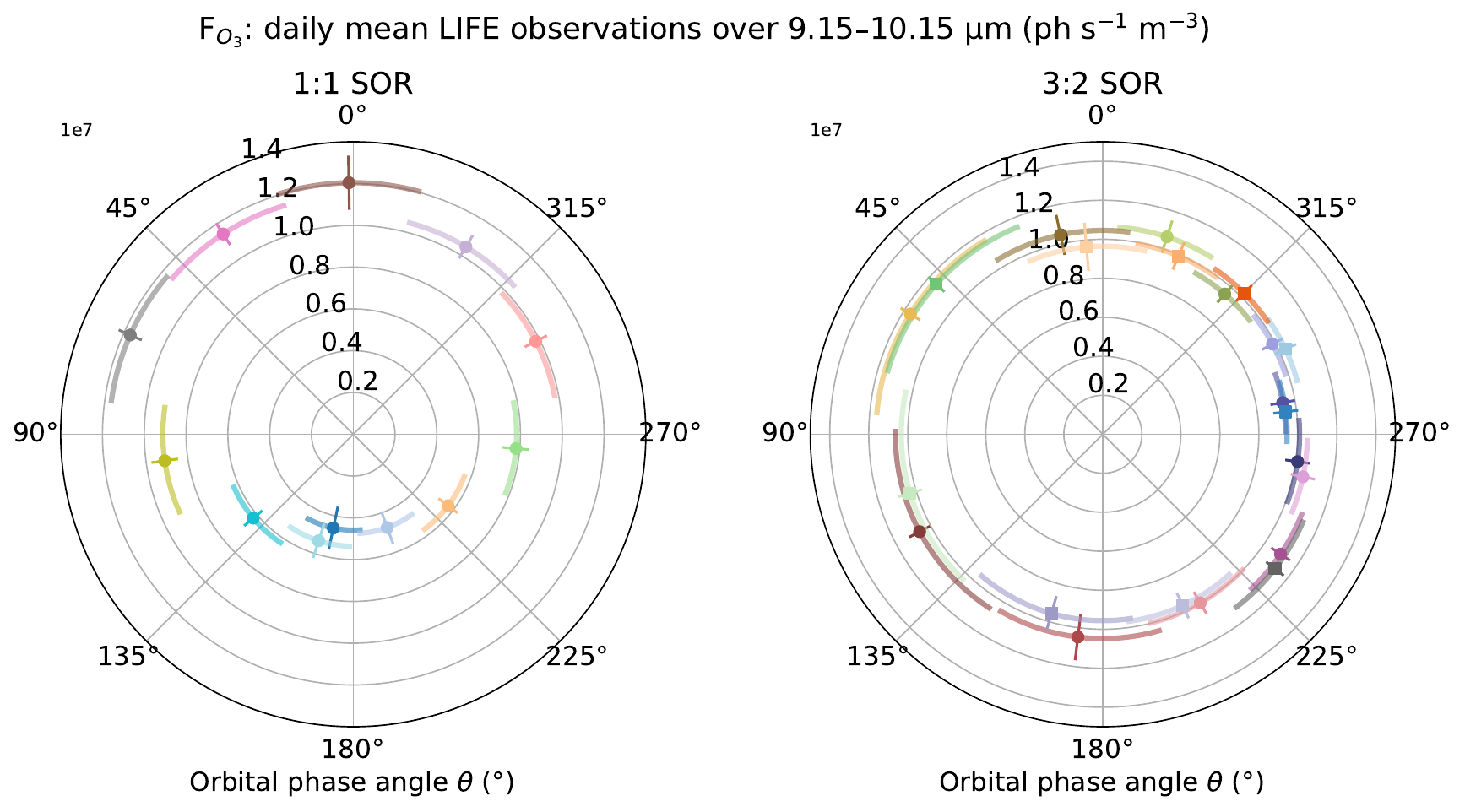}
\caption{Simulated daily observations of Proxima Centauri b for one orbit in the 1:1 SOR (left) and two orbits in the 3:2 SOR (right) with LIFE, averaged over wavelengths between 9.15--10.15~$\mu$m to probe spatial and temporal variations in the O$_3$ feature. For the 1:1 SOR, the daily averaged fluxes are spread over 32.18$^{\circ}$ corresponding to the phase angle shift per day. For the eccentric 3:2 SOR, the daily phase angle shift varies, as shown by the varying arcs of the measurements. The two orbits of the 3:2 SOR are shown by circles and squares, respectively.}
\label{fig:LIFE_FO3_daily_combined}
\end{figure*}

The distinction between both SORs becomes even more apparent when considering the temporal variations of $W_\lambda$ in Figure~\ref{fig:pcb_o3_ew}. The phase angle variations represent the temporal evolution, and colours for each SOR match across Figures~\ref{fig:pcb_emission_spectra}--\ref{fig:LIFE_FO3_daily_combined} to identify the dependence on $\theta$. For the 1:1 SOR (Figure~\ref{fig:pcb_o3_ew}a), observations of the O$_3$ band are mainly affected by the $\theta$ dependence, as the actual temporal variations of the spatial distribution are modest. For most of the orbit, the dayside hemisphere dominates the O$_3$ feature, resulting in $W_\lambda$ reaching up to 200~nm. A transition happens for $\theta$=130$^\circ$ and 233$^\circ$, where part of the O$_3$ feature is in absorption and part in emission. When the nightside hemisphere comes into view (168 and 200$^\circ$), the O$_3$ band is dominated by emission features, resulting in negative $W_\lambda$.  Clearly, substantial `seasonal variability' can be mimicked by the $\theta$-dependence of a 1:1 SOR. For the 3:2 SOR (Figure~\ref{fig:pcb_o3_ew}b), $W_\lambda$ mainly stays within 355${\pm}$10~nm, except for $\theta$ corresponding to pre-apoastron (254, 258, 262$^\circ$) with the lowest $W_\lambda$ of ${\sim}300$~nm. Since eccentricity drives these variations, they can be considered seasonal variations. Notably, the seasonal variations of $W_\lambda$ for an eccentric 3:2 SOR are similar to Earth's (348--366~nm) for most of the orbit \citep{mettler_earth_2023}. Only around apoastron, we see a much smaller $W_\lambda$ for the 3:2 SOR. Evidently, an eccentric orbit with abiotic O$_2$/O$_3$ can produce significant seasonal variations around apoastron, with a distinct dependence on $\theta$. Therefore, we need to interpret phase angle and seasonal variations in the context of the orbital configuration to robustly identify temporal variations as seasonally varying biosignatures.

% Maybe an interesting final figure here would be the statistical significance of the O$_3$ feature for LIFE observations: run all spectra through LIFESim, calculate stat significance like in FIgure 3, filter the significances between 9.15-10.15 micron, and plot these as a function of orbital phase/daily observations. Hence, 'daily phase curve'

As is evident from Sections~\ref{subsec:life_spectra} and \ref{subsec:circulation}, simulated LIFE observations will add noise to the detection and interpretation of such seasonally varying biosignatures. To understand whether LIFE can still interpret seasonal variations in atmospheric chemistry, we use the simulated daily spectra of Proxima Centauri b with LIFE and focus once again on the O$_3$ feature. We extract the daily planet fluxes between 9.15--10.15~$\mu$m to calculate F$_{O_3}$, the mean observed flux with LIFE over this wavelength region. The uncertainties in F$_{O_3}$ are calculated from the uncertainties on each spectral measurement with LIFE. Figure~\ref{fig:LIFE_FO3_daily_combined} shows the resulting evolution of F$_{O_3}$ (radial) with phase angle (polar), for the 1:1 SOR (left) and 3:2 SOR (right). The radial error represents the uncertainty in the LIFE measurement and the phase angle error represents the phase angle evolution during the 24~hours of observation with LIFE. The planet flux variations with phase angle for the 1:1 SOR are clearly observable with LIFE, as LIFE detects planet fluxes up to three times larger for $\theta{=}0^{\circ}$ compared to $\theta{=}180^{\circ}$. In between, F$_{O_3}$ decreases with $\theta$ for $\theta{<}180^{\circ}$ and increases for $\theta{>}180^{\circ}$. For a 1:1 SOR, the varying observed planet flux implies that observations at multiple phase angles may be required for a conclusive (non)-detection of a biosignature gas. 

Unsurprisingly, daily observations of the 3:2 SOR with LIFE show considerably smaller variation with phase angle (right panel of Figure~\ref{fig:LIFE_FO3_daily_combined}). Due to its eccentric orbit, the phase angle coverage per day is no longer constant, explaining the varying arc lengths. Furthermore, we show both orbits in the plot, separated by markers (circles and squares). The majority of simulated daily measurements fall within the range of 0.8--1.2 photons~s$^{-1}$~m$^{-3}$ and no clear pattern with phase angle is identifiable. Hence, the trends for F$_{O_3}$ with $\theta$ are reminiscent of those for $W_\lambda$ in Figure~\ref{fig:pcb_o3_ew} for both SORs. Evidently, LIFE can disentangle both resonances and will provide an unprecedented window into the atmospheres of nearby exoplanets. For the most accessible exoplanets, this amounts to potential daily characterisation of the 4D physical and chemical state of their atmospheres. 

\section{Discussion}\label{sec:discussion}

\subsection{Prospects for atmospheric characterisation with LIFE}
The synthetic observations of Proxima Centauri b demonstrate that MIR spectroscopy with LIFE confidently detects atmospheric molecules such as CO$_2$, H$_2$O, and O$_3$ on rocky exoplanets, agreeing with \citet{LIFE3, LIFE5}. Molecular absorption features and continuum emission enable differentiation between different SORs, as shown here for 1:1 and 3:2 SORs, given an appropriate phase angle or temporal coverage of the planetary orbits. This spectral variability is a powerful tool to infer planetary rotation states and atmospheric dynamics. While O$_3$ remains a key biosignature candidate, its interpretation is complicated by plausible abiotic production pathways on M-dwarf planets through photochemical processes \citep[e.g.,][]{hu2012photochemistry, domagal2014abiotic, tian2014high, harman2015abiotic}, underscoring the need for contextual understanding. The relation between O$_3$ and O$_2$ depends on their respective abundances and varies with stellar type and abundances of other trace gases like N$_2$O and CH$_4$ \citep{kozakis_is_2022, 2025A&A...699A.247K, 2025A&A...701A.254K}. Spatial and temporal variations in O$_3$ abundances and spectral features further complicate the derivation of O$_2$ abundances from O$_3$, and necessitate a comprehensive understanding of the planetary environment driving O$_3$ variability. The combined detection of O$_3$ and CH$_4$ in exoplanet spectra presents a more robust biosignature \citep[e.g.,][]{schwieterman_exoplanet_2018} and future work should assess the coupled spatial and temporal evolution of O$_3$ and CH$_4$ and the resulting spectral variability.

Our results demonstrate LIFE's strengths in resolving spatial and temporal variations in atmospheric composition and temperature. Firstly, LIFE can reveal circulation patterns: planets in synchronous rotation (1:1 SOR) feature notable phase-dependent flux changes related to the eastward advection of the substellar cloud deck \citep{carone_connecting_2015, haqq-misra_demarcating_2018, sergeev_bistability_2022}. These are detectable by LIFE by comparing planet fluxes at phase angles around 90$^{\circ}$ to fluxes around 270$^{\circ}$. Our analysis shows that variations in the equivalent width of O$_3$ features (W$_\lambda$) for the 3:2 SOR are similar to the seasonal variability on Earth \citep[${\sim}$5\%;][]{mettler_earth_2023}. Exoplanets in 1:1 SOR display much larger variability, including transitions from absorption to emission in O$_3$ features when the nightside comes into view. The disappearance of O$_3$ features at nightside phases agrees with Earth observations \citep{mettler_earth_2023} showing O$_3$ and CO$_2$ detection challenges for viewing angles centred on the polar regions. \citet{hearty_mid-infrared_2009} note that the polar viewing angles can also produce emission features in O$_3$ and water vapour arising due to temperature inversions, similar to the nightside viewing angles for the 1:1 SOR here. This clearly illustrates that spatial variability is significant for synchronous exoplanets, and phase-resolved observations can therefore reveal the 3D thermal structure. We find that phase angle degeneracies between different SORs persist without full phase coverage, necessitating phase-resolved observations rather than single snapshot spectra.

Due to the coupled impact of temperature, cloud, and ozone distributions, the interpretation of O$_3$ variability is inherently tied to correctly interpreting the temperature and cloud distributions and necessitates a 4D modelling context. Since $W_\lambda$ (Figure~\ref{fig:pcb_o3_ew}) calculates the strength of the O$_3$ feature relative to the continuum signal, it contains information on both the O$_3$ distribution and the atmospheric structure. The predicted phase evolution and periodicity from 4D CCMs can further limit the number of possible degeneracies. Previous studies have shown that vertically constant temperature profiles, gaseous abundances, and non-patchy cloud profiles in retrievals bias the retrieved posteriors \citep[in the context of LIFE, see][]{alei_large_2022, mettler_earth_2024, konrad2024pursuing}.Including physically motivated profiles for these parameters enables accurate retrievals of P–T structure, H$_2$O and cloud profiles, and gaseous abundances from Earth's thermal emission spectrum \citep{konrad2024pursuing}, and will be essential for capturing the spatial and temporal variability predicted in this study. Future work should conduct a dedicated retrieval of phase-resolved synthetic LIFE spectra for exoplanets in SORs, including such physically motivated vertical profiles. 

The interpretation of spatial and temporal variations is also crucial for inferring potentially habitable environments. Planets in 3:2 SOR show higher planet fluxes and moderated temporal variability due to eccentricity-driven heating and weakened dayside cloud feedback \citep{colose_effects_2021, yang2013stabilizing}. Conversely, planets in 1:1 SOR maintain more static physical and chemical environments, with stronger spatial gradients. On planets in 1:1 SOR, habitability is plausible in the terminator regions \citep{shields_habitability_2016, lobo2023terminator}, but stronger cycling of UV radiation, temperature, and wet-dry phases thought critical for life's origin and prevalence \citep{pearce2017origin, del_genio_habitable_2019} appear more prominent for 3:2 SORs. Water vapour spectral features and continuum variations modulated by clouds can serve as proxies for these surface wet-dry cycles.

We note that the reference architecture of LIFE has evolved from the setup assumed in this study (Table~\ref{tab:lifesim}), most notably in terms of sensitivity (aperture diameter) and spectral resolution. Recent simulations of the optimal LIFE configuration point towards larger aperture sizes (F. Dannert et al., private comm.). For our analysis, this implies that the quality of spectra as presented in Figures~\ref{fig:LIFE_spectra_all}, ~\ref{fig:pcb_spectra_circ}, and ~\ref{fig:LIFE_FO3_daily_combined} could be achieved with shorter observation times, or, conversely, that maintaining 24-hour integrations would yield even higher sensitivity for daily physical and chemical characterisation of Proxima Centauri b. Moreover, the planet's small angular separation (${\sim}$60 mas) ensures that the reduction in field-of-view taper associated with a larger aperture will not affect the observations (Hansen et al., in prep.). Finally, an increased spectral resolution of R=100 \citep{konrad2024pursuing} would further enhance spectral characterisation and can, if necessary, be downsampled to R=50. 

The synthetic observations we generated using LIFE\textsc{sim} consider astrophysical noise. In a realistic observing campaign, instrument systematic noise can induce offsets in the measured spectra, e.g. through temporal variability of the null floor from instrument perturbations, which in turn can mask the effects of atmospheric variability. Recent LIFE‑focused studies have begun to quantify what this means for the instrument design of LIFE \citep{2025AJ....170..193D, 2025AJ....170..227H, rutten_2025}, demonstrating that post‑processing techniques such as data whitening can substantially mitigate these additional noise contributions \citep{2025AJ....170..227H}. We therefore expect that, while systematic null‑floor variability may reduce absolute sensitivity to some degree, the impact on the inter-phase differences and diagnostic comparisons presented in this study can be mitigated.

\subsection{Limitations to the model configuration}
The predicted spectral differences between SORs depend on the assumed surface conditions, atmospheric composition and pressure, and atmospheric circulation regimes. \citet{fujii_probing_2025} analyse spatial gradients and their effects on MIR spectra of rocky exoplanets in 1:1 SOR for diverse surface scenarios. Global slab oceans tend to homogenise phase-dependent continuum emission, whereas atmospheres without oceans exhibit enhanced phase variability despite the presence of water vapour \citep{fujii_probing_2025}. Since our simulations concern global slab oceans, the predicted MIR variability might be enhanced in scenarios with more land coverage. On the other hand, \citet{del_genio_habitable_2019} show that global dynamic oceans (i.e., including ocean heat transport) homogenise temperature distributions. Nevertheless, zonal asymmetries in geopotential height for dynamic ocean scenarios suggest that circulation-driven ozone asymmetries persist. Combining both arguments, an assumed global slab ocean may be a middle case in terms of MIR variability, with enhanced land coverage enhancing spectral variability and global dynamic oceans reducing it. The inclusion of orography affects atmospheric circulation and thus the ozone distribution \citep[][]{bhongade_asymmetries_2024}. Future work should elucidate the robustness of predicted O$_3$, H$_2$O, and CO$_2$ variability for varying land cover and dynamic oceans, e.g. using an efficient two-layer dynamical slab ocean model \citep{bhatnagar2025fast}. 

%circulation regime 
Exoplanets in 1:1 SOR exhibit specific circulation regimes depending on the orbital period \citep{carone_connecting_2015, noda_circulation_2017, haqq-misra_demarcating_2018}. Moreover, for planets close to regime transitions, model parameterisations can flip the regime \citep{sergeev_atmospheric_2020}, or planets can exhibit bistability \citep{sergeev_bistability_2022}. The circulation regime is crucial to the 4D distributions of, for example, temperature, humidity, and photochemical species. However, the atmosphere of Proxima Centauri b is comparatively stable against regime changes. \citet{carone2018stratosphere} show in their figure 1 that Proxima Centauri b falls in the weak superrotation or single jet regime. Compared to this, TRAPPIST-1 e falls on the boundary between two circulation regimes, making it particularly susceptible to regime changes \citep{sergeev_bistability_2022}. For Proxima b, a different convection parameterisation scheme also does not change the circulation regime \citep{sergeev_atmospheric_2020}. Nevertheless, the sensitivity of the simulated distributions can be further evaluated against diverse landmasses \citep{lewis2018influence} and tidal heating \citep{colose_effects_2021}.

Our simulations assume a 1 bar pre-industrial Earth-like atmosphere of N$_2$, O$_2$, and CO$_2$ as a fiducial case. However, we note that the results presented will be affected by different atmospheric compositions and pressures. \citet{turbet_habitability_2016} explore the effects of diverse atmospheric scenarios on the climate of Proxima b. They show that the zonal variations in surface temperature for a 1:1 SOR, including the high-latitude gyres as the coldest regions on the planet, hold for atmospheres with and without water vapour as well as pressure variations of 1 to 20~bar. Nevertheless, varying gyre positions and prominence potentially affect the zonal variations. The 3:2 SOR, in this case without eccentricity, consistently displays latitudinal variations, but enhanced CO$_2$ abundances lower the temperature contrasts. Although investigations of ozone photochemistry are mostly limited to pre-industrial Earth-like conditions, investigations of the ozone distributions on exoplanets in 1:1 SOR with lower O$_2$ levels (down to 1\% or 0.1\% present atmospheric level) show that the zonal variations persist \citep{braam_stratospheric_2023, cooke2023degenerate}.

\citet{mak20243d} use the UM to simulate the atmosphere of TRAPPIST-1 e in 1:1 SOR, with varying CO$_2$/CH$_4$ partial pressures as well as prescribed haze layers corresponding to the CH$_4$ levels, in a background of N$_2$. For increasing CO$_2$ levels, the circulation regime and zonal temperature variations persist, although temperature contrasts decrease. Varying CH$_4$ levels can change the circulation regime, substantially affecting the zonal temperature variations, but higher CH$_4$ abundances cool the surface. When hazes are included, further surface cooling is predicted, along with enhanced zonal temperature variations \citep{mak20243d}. As noted by \citet{braam_stratospheric_2023}, the circulation-driven chemical distributions likely hold for any chemical compounds resulting from stratospheric photochemistry. Therefore, photochemical hazes or other products from methane photodissociation may show similar spatial variations. The 4D aspects of this photochemistry need further study, including their MIR observability and variability.

\subsection{Abiotic scenarios for seasonally varying biosignatures}
Seasonally varying biosignatures are considered among the strongest indicators of life \citep{olson2018atmospheric, schwieterman_exoplanet_2018}. Clouds modulate spectral intensities and variability amplitudes \citep{tinetti_detectability_2006, des_marais_remote_2002, kitzmann_clouds_2011}, but O$_3$ location above cloud layers mitigates this effect. However, O$_3$ features tend to saturate, limiting precise abundance constraints despite phase-dependent flux changes. The variations in W$_\lambda$ of O$_3$ translate into small and big variations for predicted LIFE spectra of the 3:2 and 1:1 SOR, and such variations would be further enhanced if landmasses were included \citep{mettler_earth_2023}. Within typical error bars, the spatial and temporal variations in O$_3$ are detectable, mimicking signals due to seasonally varying biosignatures \citep{olson2018atmospheric, schwieterman_exoplanet_2018}. As noted before, abiotic pathways to O$_3$ production exist on planets around M-dwarfs \citep[e.g.,][]{hu2012photochemistry, domagal2014abiotic, tian2014high, harman2015abiotic}. Therefore, such `abiotic' spatial and temporal variations due to orbital geometry or internal atmospheric variability \citep[see also][]{cohen2023traveling, cooke_variability_2023, braam_earth-like_2025} are possible for tidally locked planets and potentially appear as false positives in LIFE spectra. Interpreting the spectra alongside 4D model predictions can circumvent the ambiguity by providing the expected periodicity of abiotic variability.

\subsection{Observational recommendations for LIFE}\label{sec:disc_LIFE_observing}
We obtain little insight into key biosignatures and their variability in the 4--6.5~$\mu$m spectral range, given the predicted noise for LIFE observations of Proxima Centauri b. This is partly due to baseline optimisation focused on longer wavelengths and the total planetary flux. Alternatively, LIFE observations can change the baseline to optimise for shorter wavelengths, both for Proxima Centauri b or other targets (not shown). Wavelengths ${>}$16~$\mu$m, on the other hand, are affected by water vapour features with statistically significant variability for the 1:1 SOR and should be prioritised. Enhanced spectral resolution near the 9.6~$\mu$m O$_3$ band appears unnecessary given the detectable variability effects in O$_3$ for the 1:1 SOR. The predicted signal-to-noise allows for confidently detecting or ruling out O$_3$ and water vapour variability around Proxima Centauri b, provided observations are taken at multiple phase angles and not as snapshot spectra. The detection of variability and distinction between SORs is, in essence, a comparison of timescales. Observation times must be balanced against variability timescales: longer integrations will improve S/N but can lead to unresolved variability by averaging over phase angles, potentially hiding dynamic features. Careful observation planning should therefore include a determination of optimal phase angle coverage given dynamical and (photo)chemical timescales.

Studying the phase dependence of exoplanetary signals observed with LIFE thus also plays an important role in the ongoing definition of the observing strategy. In current yield studies such as \cite{LIFE1, LIFE6}, planets are assumed as `static' with randomised orbital phase angles, i.e. not significantly changing throughout the observation. Such snapshot observations at random phase angles, used to identify potential rocky planet candidates during the detection phase of LIFE, could lead to confusion about the nature of the planet (e.g., missing O$_3$ signature, how to fit temperature/blackbody). The results presented here can be used to optimise and refine the detection strategy by guiding the timing of multiple visits for confirming detected planet candidates and informing the transition from the detection to the characterisation phase (e.g., by excluding planets exhibiting large phase variations). Most of LIFE's characterisation observations \cite[cf.][]{LIFE3, LIFE5} require long integration time and, depending on planetary spin rates and tidal locking, need to take into account the 4D structure of this problem, i.e. that they average over changing phase angles.

Time integrations for LIFE observations of any planet include unresolved spatial and temporal variations, but the orbital period and distance from Earth make Proxima Centauri b a golden target for LIFE \citep{LIFE12}. By combining daily average climate model output and daily observations with LIFE, we demonstrate LIFE's potential to provide daily characterisation of the physical and chemical state of the atmosphere of Proxima Centauri b. Planets on shorter orbital periods face larger daily phase angle changes and therefore lose spatial and temporal resolution in LIFE observations. Temperate planets on longer orbital periods around brighter stars may still be tidally locked into SORs within their lifetime \citep{barnes_tidal_2017}. We perform initial tests that suggest that analogous variability in MIR spectra is accessible around brighter stars, but requires enhanced observational time (see Section~\ref{app:brighter_stars}). As such, there will be a substantial number of exoplanetary systems within ${\sim}$5-7 parsecs accessible for LIFE characterisation in a similar manner. For instance, \citet{LIFE12} showed that LIFE has an expected yield of about a dozen HZ planets in late-type systems and a handful in FGK-host systems within 6~pcs. For planets on longer periods around brighter stars, longer observations may cover phase angle variations similar to those in our study, thus providing similar spatial and temporal resolution. Nevertheless, the associated differences in host star spectral type, orbital configuration, atmospheric dynamics, and atmospheric chemistry will modulate the signals observed by LIFE, motivating broader modelling efforts of 4D atmospheric chemistry for planets around K- and G-type stars with varying orbital periods. 

In this study, we used an intermediate inclination (${\sim}70^{\circ}$) to determine the effects of spatial and temporal variations on emission spectra with LIFE, reflecting estimates for Proxima Centauri b and other 3D climate modelling studies \citep{turbet_habitability_2016, 2017AJ....153...52K, boutle_exploring_2017, braam_earth-like_2025}. For this inclination, Proxima Centauri b stays within the optimal modulation efficiency range of LIFE. However, other inclinations are possible, where targets like Proxima Centauri b temporarily leave the optimal modulation efficiency range. Additionally, this would change the planetary mass (and radius) of Proxima Centauri b through the $M\sin (i) {=} 1.27~M_{\oplus}$ dependence. A (significantly) smaller inclination modulates the phase variability, but previous work has shown that emission spectra are less sensitive than reflection spectra \citep{turbet_habitability_2016, boutle_exploring_2017}. Future work should investigate inclination variations to further constrain LIFE's ability to characterise spatial and temporal variations on terrestrial exoplanets and to determine whether temporarily leaving the optimal modulation efficiency range of LIFE affects this characterisation. We note that inferences on planetary atmosphere and surface conditions from spectra depend on model assumptions \citep{paradise2022fundamental, fauchez_trappist-1_2022}, motivating further development of coupled 4D climate-chemistry models and retrieval methods considering spatial-temporal variability.

\section{Conclusions}\label{sec:conclusion}
This study combines comprehensive 4D climate-chemistry modelling with synthetic mid-infrared observations to investigate LIFE's capability to characterise spatial and temporal variations in the atmosphere of Proxima Centauri b and similar nearby exoplanets. Our key findings are as follows:
\begin{itemize}
   \item  LIFE enables daily, mid-infrared characterisation of atmospheric composition and variability on Proxima Centauri b, confidently distinguishing between different spin-orbit resonance scenarios through observable differences in O$_3$, water vapour, and temperature signatures.
   \item  The synchronous (1:1) spin-orbit resonance produces strong hemispheric chemical and thermal contrasts, with detectable phase-dependent O$_3$ features and temperature inversions on the nightside, while the eccentric (3:2) case yields a more homogeneous atmosphere and enhanced global flux due to variable stellar irradiation.
   \item  Simulated LIFE observations demonstrate that 4D (spatial and temporal) atmospheric variability in rocky exoplanets can be resolved, allowing discrimination of dynamical and chemical regimes and providing detailed insight into atmospheric circulation mechanisms such as the advection of cloud decks and O$_3$ accumulation.
   \item  The study highlights that phase-dependent molecular features can act as false positives for seasonal biosignatures, in the case of abiotic O$_2$/O$_3$ build-up, emphasising the need for robust interpretation strategies that account for orbital and planetary context when assessing biosignatures on M-dwarf planets.
   \item The results advocate further 4D modelling efforts and optimised observing strategies for LIFE and similar missions, suggesting that dozens of exoplanets within 7 parsecs may be accessible for 4D daily characterisation, advancing the search for biosignatures and our understanding of terrestrial planet atmospheres.
\end{itemize}

\begin{acknowledgements}
We thank Thomas Fauchez, Vincent Kofman, and Geronimo Villanueva for advice on generating emission spectra from a 3D model using PSG. We thank Andrea Fortier, Felix Dannert, Philipp Huber and Jonah Hansen for helpful discussions about LIFE\textsc{sim}, baselines and instrument requirements in general. We thank Felix Dannert, Eleonora Alei, and Sascha Quanz for valuable feedback on the manuscript. We are grateful to the anonymous reviewer whose comments helped to
significantly improve the manuscript.

MB appreciates support from a CSH Fellowship. For the CCM simulations, we gratefully acknowledge the use of the MONSooN2 system, a collaborative facility supplied under the Joint Weather and Climate Research Programme, a strategic partnership between the UK Met Office and the Natural Environment Research Council. Our simulations were performed as part of the project space “Using UKCA to investigate atmospheric composition on extra-solar planets (ExoChem).”
\end{acknowledgements}

\bibliographystyle{aa}
\bibliography{references_v2}

@article{yang2013stabilizing,
	title        = {Stabilizing cloud feedback dramatically expands the habitable zone of tidally locked planets},
	author       = {Yang, Jun and Cowan, Nicolas B and Abbot, Dorian S},
	year         = 2013,
	journal      = {The Astrophysical Journal Letters},
	publisher    = {IOP Publishing},
	volume       = 771,
	number       = 2,
	pages        = {L45}
}

@article{luo2023coupled,
	title        = {Coupled atmospheric chemistry, radiation, and dynamics of an exoplanet generate self-sustained oscillations},
	author       = {Luo, Yangcheng and Hu, Yongyun and Yang, Jun and Zhang, Michael and Yung, Yuk L},
	year         = 2023,
	journal      = {Proceedings of the National Academy of Sciences},
	publisher    = {National Academy of Sciences},
	volume       = 120,
	number       = 51,
	pages        = {e2309312120}
}

@article{hu2012photochemistry,
	title        = {Photochemistry in terrestrial exoplanet atmospheres. I. Photochemistry model and benchmark cases},
	author       = {Hu, Renyu and Seager, Sara and Bains, William},
	year         = 2012,
	journal      = {The Astrophysical Journal},
	publisher    = {IOP Publishing},
	volume       = 761,
	number       = 2,
	pages        = 166
}

@article{tian2014high,
	title        = {High stellar FUV/NUV ratio and oxygen contents in the atmospheres of potentially habitable planets},
	author       = {Tian, Feng and France, Kevin and Linsky, Jeffrey L and Mauas, Pablo JD and Vieytes, Mariela C},
	year         = 2014,
	journal      = {Earth and Planetary Science Letters},
	publisher    = {Elsevier},
	volume       = 385,
	pages        = {22--27}
}

@article{domagal2014abiotic,
	title        = {Abiotic ozone and oxygen in atmospheres similar to prebiotic Earth},
	author       = {Domagal-Goldman, Shawn D and Segura, Ant{\'\i}gona and Claire, Mark W and Robinson, Tyler D and Meadows, Victoria S},
	year         = 2014,
	journal      = {The Astrophysical Journal},
	publisher    = {IOP Publishing},
	volume       = 792,
	number       = 2,
	pages        = 90
}

@article{harman2015abiotic,
	title        = {Abiotic O2 levels on planets around F, G, K, and M stars: possible false positives for life?},
	author       = {Harman, CE and Schwieterman, EW and Schottelkotte, James C and Kasting, JF},
	year         = 2015,
	journal      = {The Astrophysical Journal},
	publisher    = {IOP Publishing},
	volume       = 812,
	number       = 2,
	pages        = 137
}

@article{cooke2024lethal,
	title        = {Lethal surface ozone concentrations are possible on habitable zone exoplanets},
	author       = {Cooke, GJ and Marsh, DR and Walsh, C and Sainsbury-Martinez, F},
	year         = 2024,
	journal      = {The Planetary Science Journal},
	publisher    = {IOP Publishing},
	volume       = 5,
	number       = 7,
	pages        = 168
}

@article{2020ApJ...893..140G,
	title        = {{The Impact of Planetary Rotation Rate on the Reflectance and Thermal Emission Spectrum of Terrestrial Exoplanets around Sunlike Stars}},
	author       = {{Guzewich}, Scott D. and {Lustig-Yaeger}, Jacob and {Davis}, Christopher Evan and {Kopparapu}, Ravi Kumar and {Way}, Michael J. and {Meadows}, Victoria S.},
	year         = 2020,
	month        = apr,
	journal      = {\apj},
	volume       = 893,
	number       = 2,
	pages        = 140,
	doi          = {10.3847/1538-4357/ab83ec},
	eid          = 140,
	archiveprefix = {arXiv},
	eprint       = {2002.02549},
	primaryclass = {astro-ph.EP},
	adsurl       = {https://ui.adsabs.harvard.edu/abs/2020ApJ...893..140G}
}

@article{braam_earth-like_2025,
	title        = {Earth-like {Exoplanets} in {Spin}–{Orbit} {Resonances}: {Climate} {Dynamics}, {3D} {Atmospheric} {Chemistry}, and {Observational} {Signatures}},
	shorttitle   = {Earth-like {Exoplanets} in {Spin}–{Orbit} {Resonances}},
	author       = {Braam, Marrick and Palmer, Paul I. and Decin, Leen and Mayne, Nathan J. and Manners, James and Rugheimer, Sarah},
	year         = 2025,
	month        = jan,
	journal      = {The Planetary Science Journal},
	volume       = 6,
	number       = 1,
	pages        = 5,
	doi          = {10.3847/PSJ/ad9565},
	issn         = {2632-3338},
	url          = {https://iopscience.iop.org/article/10.3847/PSJ/ad9565/meta},
	urldate      = {2025-02-26},
	note         = {Publisher: IOP Publishing},
	language     = {en}
}

@article{fauchez_trappist-1_2022,
	title        = {The {TRAPPIST}-1 {Habitable} {Atmosphere} {Intercomparison} ({THAI}). {III}. {Simulated} {Observables}—the {Return} of the {Spectrum}},
	author       = {Fauchez, Thomas J. and Villanueva, Geronimo L. and Sergeev, Denis E. and Turbet, Martin and Boutle, Ian A. and Tsigaridis, Kostas and Way, Michael J. and Wolf, Eric T. and Domagal-Goldman, Shawn D. and Forget, François and Haqq-Misra, Jacob and Kopparapu, Ravi K. and Manners, James and Mayne, Nathan J.},
	year         = 2022,
	month        = sep,
	journal      = {The Planetary Science Journal},
	volume       = 3,
	number       = 9,
	pages        = 213,
	doi          = {10.3847/PSJ/ac6cf1},
	issn         = {2632-3338},
	url          = {https://iopscience.iop.org/article/10.3847/PSJ/ac6cf1/meta},
	urldate      = {2022-09-22},
	note         = {Publisher: IOP Publishing},
	language     = {en}
}

@article{kofman_pale_2024,
	title        = {The {Pale} {Blue} {Dot}: {Using} the {Planetary} {Spectrum} {Generator} to {Simulate} {Signals} from {Hyperrealistic} {Exo}-{Earths}},
	shorttitle   = {The {Pale} {Blue} {Dot}},
	author       = {Kofman, Vincent and Villanueva, Geronimo Luis and Fauchez, Thomas J. and Mandell, Avi M. and Johnson, Ted M. and Payne, Allison and Latouf, Natasha and Kelkar, Soumil},
	year         = 2024,
	month        = sep,
	journal      = {The Planetary Science Journal},
	volume       = 5,
	number       = 9,
	pages        = 197,
	doi          = {10.3847/PSJ/ad6448},
	issn         = {2632-3338},
	url          = {https://iopscience.iop.org/article/10.3847/PSJ/ad6448/meta},
	urldate      = {2025-02-26},
	note         = {Publisher: IOP Publishing},
	language     = {en}
}

@misc{manners_socrates_2021,
	title        = {{SOCRATES} ({Suite} {Of} {Community} {RAdiative} {Transfer} codes based on {Edwards} and {Slingo}) {Technical} {Guide}},
	author       = {Manners, J. and Edwards, J. M. and Hill, P. and Thelen, J.-C.},
	year         = 2021,
	publisher    = {Met Office, UK},
	url          = {https://code.metoffice.gov.uk/trac/socrates}
}

@book{smart_text-book_1944,
	title        = {Text-book on {Spherical} {Astronomy}},
	author       = {Smart, William Marshall},
	year         = 1944,
	publisher    = {The University Press},
	note         = {Google-Books-ID: SnPvAAAAMAAJ},
	language     = {en}
}

@article{mueller_equation_1995,
	title        = {Equation of {Time} - {Problem} in {Astronomy}},
	author       = {Mueller, M.},
	year         = 1995,
	journal      = {Acta Physica Polonica A 88 Supplement},
	volume       = {S-49}
}

@article{goldreich_spin-orbit_1966,
	title        = {Spin-orbit coupling in the solar system},
	author       = {Goldreich, Peter and Peale, Stanton},
	year         = 1966,
	month        = aug,
	journal      = {The Astronomical Journal},
	volume       = 71,
	pages        = 425,
	doi          = {10.1086/109947},
	issn         = {0004-6256},
	url          = {https://ui.adsabs.harvard.edu/abs/1966AJ.....71..425G},
	urldate      = {2022-08-30},
	note         = {ADS Bibcode: 1966AJ.....71..425G}
}

@article{dobrovolskis_spin_2007,
	title        = {Spin states and climates of eccentric exoplanets},
	author       = {Dobrovolskis, Anthony R.},
	year         = 2007,
	month        = dec,
	journal      = {Icarus},
	volume       = 192,
	number       = 1,
	pages        = {1--23},
	doi          = {10.1016/j.icarus.2007.07.005},
	issn         = {0019-1035},
	url          = {https://www.sciencedirect.com/science/article/pii/S0019103507003077},
	urldate      = {2022-10-12},
	language     = {en}
}

@article{del_genio_habitable_2019,
	title        = {Habitable {Climate} {Scenarios} for {Proxima} {Centauri} b with a {Dynamic} {Ocean}},
	author       = {Del Genio, Anthony D. and Way, Michael J. and Amundsen, David S. and Aleinov, Igor and Kelley, Maxwell and Kiang, Nancy Y. and Clune, Thomas L.},
	year         = 2019,
	month        = jan,
	journal      = {Astrobiology},
	volume       = 19,
	number       = 1,
	pages        = {99--125},
	doi          = {10.1089/ast.2017.1760},
	issn         = {1531-1074},
	url          = {https://www.liebertpub.com/doi/full/10.1089/ast.2017.1760},
	urldate      = {2022-04-20},
	note         = {Publisher: Mary Ann Liebert, Inc., publishers}
}

@article{turbet_habitability_2016,
	title        = {The habitability of {Proxima} {Centauri} b: {II}. {Possible} climates and observability},
	shorttitle   = {The habitability of {Proxima} {Centauri} b},
	author       = {Turbet, Martin and Leconte, Jérémy and Selsis, Franck and Bolmont, Emeline and Forget, François and Ribas, Ignasi and Raymond, Sean N. and Anglada-Escudé, Guillem},
	year         = 2016,
	month        = dec,
	journal      = {Astronomy \& Astrophysics},
	volume       = 596,
	pages        = {A112},
	doi          = {10.1051/0004-6361/201629577},
	issn         = {0004-6361, 1432-0746},
	url          = {http://www.aanda.org/10.1051/0004-6361/201629577},
	urldate      = {2020-11-06},
	language     = {en}
}

@article{anglada-escude_terrestrial_2016,
	title        = {A terrestrial planet candidate in a temperate orbit around {Proxima} {Centauri}},
	author       = {Anglada-Escudé, Guillem and Amado, Pedro J. and Barnes, John and Berdiñas, Zaira M. and Butler, R. Paul and Coleman, Gavin A. L. and de la Cueva, Ignacio and Dreizler, Stefan and Endl, Michael and Giesers, Benjamin and Jeffers, Sandra V. and Jenkins, James S. and Jones, Hugh R. A. and Kiraga, Marcin and Kürster, Martin and López-González, Marίa J. and Marvin, Christopher J. and Morales, Nicolás and Morin, Julien and Nelson, Richard P. and Ortiz, José L. and Ofir, Aviv and Paardekooper, Sijme-Jan and Reiners, Ansgar and Rodríguez, Eloy and Rodrίguez-López, Cristina and Sarmiento, Luis F. and Strachan, John P. and Tsapras, Yiannis and Tuomi, Mikko and Zechmeister, Mathias},
	year         = 2016,
	month        = aug,
	journal      = {Nature},
	volume       = 536,
	number       = 7617,
	pages        = {437--440},
	doi          = {10.1038/nature19106},
	issn         = {1476-4687},
	url          = {https://www.nature.com/articles/nature19106},
	urldate      = {2022-05-13},
	copyright    = {2016 Macmillan Publishers Limited, part of Springer Nature. All rights reserved.},
	note         = {Number: 7617 Publisher: Nature Publishing Group},
	language     = {en}
}

@article{braam_lightning-induced_2022,
	title        = {Lightning-induced chemistry on tidally-locked {Earth}-like exoplanets},
	author       = {Braam, Marrick and Palmer, Paul I and Decin, Leen and Ridgway, Robert J and Zamyatina, Maria and Mayne, Nathan J and Sergeev, Denis E and Abraham, N Luke},
	year         = 2022,
	month        = dec,
	journal      = {Monthly Notices of the Royal Astronomical Society},
	volume       = 517,
	number       = 2,
	pages        = {2383--2402},
	doi          = {10.1093/mnras/stac2722},
	issn         = {0035-8711},
	url          = {https://doi.org/10.1093/mnras/stac2722},
	urldate      = {2023-04-11}
}

@article{edwards_studies_1996,
	title        = {Studies with a flexible new radiation code. {I}: {Choosing} a configuration for a large-scale model},
	shorttitle   = {Studies with a flexible new radiation code. {I}},
	author       = {Edwards, J. M. and Slingo, A.},
	year         = 1996,
	journal      = {Quarterly Journal of the Royal Meteorological Society},
	volume       = 122,
	number       = 531,
	pages        = {689--719},
	doi          = {https://doi.org/10.1002/qj.49712253107},
	issn         = {1477-870X},
	url          = {https://rmets.onlinelibrary.wiley.com/doi/abs/10.1002/qj.49712253107},
	urldate      = {2021-05-26},
	language     = {en}
}

@article{archibald_description_2020,
	title        = {Description and evaluation of the {UKCA} stratosphere–troposphere chemistry scheme ({StratTrop} vn 1.0) implemented in {UKESM1}},
	author       = {Archibald, Alexander T. and O'Connor, Fiona M. and Abraham, Nathan Luke and Archer-Nicholls, Scott and Chipperfield, Martyn P. and Dalvi, Mohit and Folberth, Gerd A. and Dennison, Fraser and Dhomse, Sandip S. and Griffiths, Paul T. and Hardacre, Catherine and Hewitt, Alan J. and Hill, Richard S. and Johnson, Colin E. and Keeble, James and Köhler, Marcus O. and Morgenstern, Olaf and Mulcahy, Jane P. and Ordóñez, Carlos and Pope, Richard J. and Rumbold, Steven T. and Russo, Maria R. and Savage, Nicholas H. and Sellar, Alistair and Stringer, Marc and Turnock, Steven T. and Wild, Oliver and Zeng, Guang},
	year         = 2020,
	month        = mar,
	journal      = {Geoscientific Model Development},
	volume       = 13,
	number       = 3,
	pages        = {1223--1266},
	doi          = {https://doi.org/10.5194/gmd-13-1223-2020},
	issn         = {1991-959X},
	url          = {https://gmd.copernicus.org/articles/13/1223/2020/},
	urldate      = {2021-03-01},
	note         = {Publisher: Copernicus GmbH},
	language     = {English}
}

@article{boutle_exploring_2017,
	title        = {Exploring the climate of {Proxima} {B} with the {Met} {Office} {Unified} {Model}},
	author       = {Boutle, Ian A. and Mayne, Nathan J. and Drummond, Benjamin and Manners, James and Goyal, Jayesh and Hugo Lambert, F. and Acreman, David M. and Earnshaw, Paul D.},
	year         = 2017,
	month        = may,
	journal      = {Astronomy \& Astrophysics},
	volume       = 601,
	pages        = {A120},
	doi          = {10.1051/0004-6361/201630020},
	issn         = {0004-6361, 1432-0746},
	url          = {http://www.aanda.org/10.1051/0004-6361/201630020},
	urldate      = {2020-11-06},
	language     = {en}
}

@article{mayne_using_2014,
	title        = {Using the {UM} dynamical cores to reproduce idealised 3-{D} flows},
	author       = {Mayne, N. J. and Baraffe, I. and Acreman, D. M. and Smith, C. and Wood, N. and Amundsen, D. S. and Thuburn, J. and Jackson, D. R.},
	year         = 2014,
	month        = dec,
	journal      = {Geoscientific Model Development},
	volume       = 7,
	number       = 6,
	pages        = {3059--3087},
	doi          = {10.5194/gmd-7-3059-2014},
	issn         = {1991-959X},
	url          = {https://gmd.copernicus.org/articles/7/3059/2014/},
	urldate      = {2021-10-06},
	note         = {Publisher: Copernicus GmbH},
	language     = {English}
}

@article{yates_ozone_2020,
	title        = {Ozone chemistry on tidally locked {M} dwarf planets},
	author       = {Yates, Jack S and Palmer, Paul I and Manners, James and Boutle, Ian and Kohary, Krisztian and Mayne, Nathan and Abraham, Luke},
	year         = 2020,
	month        = feb,
	journal      = {Monthly Notices of the Royal Astronomical Society},
	volume       = 492,
	number       = 2,
	pages        = {1691--1705},
	doi          = {10.1093/mnras/stz3520},
	issn         = {0035-8711, 1365-2966},
	url          = {https://academic.oup.com/mnras/article/492/2/1691/5698322},
	urldate      = {2020-11-06},
	language     = {en}
}

@article{ridgway_3d_2023,
	title        = {{3D} modelling of the impact of stellar activity on tidally locked terrestrial exoplanets: atmospheric composition and habitability},
	shorttitle   = {{3D} modelling of the impact of stellar activity on tidally locked terrestrial exoplanets},
	author       = {Ridgway, R J and Zamyatina, M and Mayne, N J and Manners, J and Lambert, F H and Braam, M and Drummond, B and Hébrard, E and Palmer, P I and Kohary, K},
	year         = 2023,
	month        = jan,
	journal      = {Monthly Notices of the Royal Astronomical Society},
	volume       = 518,
	number       = 2,
	pages        = {2472--2496},
	doi          = {10.1093/mnras/stac3105},
	issn         = {0035-8711},
	url          = {https://doi.org/10.1093/mnras/stac3105},
	urldate      = {2023-03-20}
}

@article{telford_implementation_2013,
	title        = {Implementation of the {Fast}-{JX} {Photolysis} scheme (v6.4) into the {UKCA} component of the {MetUM} chemistry-climate model (v7.3)},
	author       = {Telford, P. J. and Abraham, N. L. and Archibald, A. T. and Braesicke, P. and Dalvi, M. and Morgenstern, O. and O'Connor, F. M. and Richards, N. A. D. and Pyle, J. A.},
	year         = 2013,
	month        = feb,
	journal      = {Geoscientific Model Development},
	volume       = 6,
	number       = 1,
	pages        = {161--177},
	doi          = {10.5194/gmd-6-161-2013},
	issn         = {1991-9603},
	url          = {https://gmd.copernicus.org/articles/6/161/2013/},
	urldate      = {2021-02-26},
	language     = {en}
}

@article{walters_met_2019,
	title        = {The {Met} {Office} {Unified} {Model} {Global} {Atmosphere} 7.0/7.1 and {JULES} {Global} {Land} 7.0 configurations},
	author       = {Walters, David and Baran, Anthony J. and Boutle, Ian and Brooks, Malcolm and Earnshaw, Paul and Edwards, John and Furtado, Kalli and Hill, Peter and Lock, Adrian and Manners, James and Morcrette, Cyril and Mulcahy, Jane and Sanchez, Claudio and Smith, Chris and Stratton, Rachel and Tennant, Warren and Tomassini, Lorenzo and Van Weverberg, Kwinten and Vosper, Simon and Willett, Martin and Browse, Jo and Bushell, Andrew and Carslaw, Kenneth and Dalvi, Mohit and Essery, Richard and Gedney, Nicola and Hardiman, Steven and Johnson, Ben and Johnson, Colin and Jones, Andy and Jones, Colin and Mann, Graham and Milton, Sean and Rumbold, Heather and Sellar, Alistair and Ujiie, Masashi and Whitall, Michael and Williams, Keith and Zerroukat, Mohamed},
	year         = 2019,
	month        = may,
	journal      = {Geoscientific Model Development},
	volume       = 12,
	number       = 5,
	pages        = {1909--1963},
	doi          = {https://doi.org/10.5194/gmd-12-1909-2019},
	issn         = {1991-959X},
	url          = {https://gmd.copernicus.org/articles/12/1909/2019/},
	urldate      = {2021-03-02},
	note         = {Publisher: Copernicus GmbH},
	language     = {English}
}

@article{braam_stratospheric_2023,
	title        = {Stratospheric dayside-to-nightside circulation drives the {3D} ozone distribution on synchronously rotating rocky exoplanets},
	author       = {Braam, Marrick and Palmer, Paul I and Decin, Leen and Cohen, Maureen and Mayne, Nathan J},
	year         = 2023,
	month        = nov,
	journal      = {Monthly Notices of the Royal Astronomical Society},
	volume       = 526,
	number       = 1,
	pages        = {263--278},
	doi          = {10.1093/mnras/stad2704},
	issn         = {0035-8711},
	url          = {https://doi.org/10.1093/mnras/stad2704},
	urldate      = {2023-10-31}
}

@article{fauchez2025global,
	title        = {From global climate models (GCMs) to exoplanet spectra with the Global Emission Spectra (GlobES)},
	author       = {Fauchez, Thomas J and Villanueva, Geronimo L and Kofman, Vincent and Suissa, Gabriella and Kopparapu, Ravi K},
	year         = 2025,
	journal      = {Astronomy and Computing},
	publisher    = {Elsevier},
	pages        = 100982
}

@book{villanueva_fundamentals_2022,
	title        = {Fundamentals of the {Planetary} {Spectrum} {Generator}},
	author       = {Villanueva, Geronimo Luis and Liuzzi, Giuliano and Faggi, Sara and Protopapa, Silvia and Kofman, Vincent and Fauchez, Thomas and Stone, Shane Wesley and Mandell, Avi Max},
	year         = 2022,
	month        = jan,
	url          = {https://ui.adsabs.harvard.edu/abs/2022fpsg.book.....V},
	urldate      = {2022-04-28},
	note         = {ISBN 978-0-578-36143-7}
}

@article{villanueva_planetary_2018,
	title        = {Planetary {Spectrum} {Generator}: {An} accurate online radiative transfer suite for atmospheres, comets, small bodies and exoplanets},
	shorttitle   = {Planetary {Spectrum} {Generator}},
	author       = {Villanueva, G. L. and Smith, M. D. and Protopapa, S. and Faggi, S. and Mandell, A. M.},
	year         = 2018,
	month        = sep,
	journal      = {Journal of Quantitative Spectroscopy and Radiative Transfer},
	volume       = 217,
	pages        = {86--104},
	doi          = {10.1016/j.jqsrt.2018.05.023},
	issn         = {0022-4073},
	url          = {https://www.sciencedirect.com/science/article/pii/S0022407318301572},
	urldate      = {2022-02-25},
	language     = {en}
}

@article{mettler_earth_2023,
	title        = {Earth as an {Exoplanet}. {II}. {Earth}’s {Time}-variable {Thermal} {Emission} and {Its} {Atmospheric} {Seasonality} of {Bioindicators}},
	author       = {Mettler, Jean-Noël and Quanz, Sascha P. and Helled, Ravit and Olson, Stephanie L. and Schwieterman, Edward W.},
	year         = 2023,
	month        = apr,
	journal      = {The Astrophysical Journal},
	volume       = 946,
	number       = 2,
	pages        = 82,
	doi          = {10.3847/1538-4357/acbe3c},
	issn         = {0004-637X},
	url          = {https://dx.doi.org/10.3847/1538-4357/acbe3c},
	urldate      = {2024-02-29},
	note         = {Publisher: The American Astronomical Society},
	language     = {en}
}

@article{chen2023sporadic,
  title={Sporadic spin-orbit variations in compact multiplanet systems and their influence on exoplanet climate},
  author={Chen, Howard and Li, Gongjie and Paradise, Adiv and Kopparapu, Ravi K},
  journal={The Astrophysical Journal Letters},
  volume={946},
  number={2},
  pages={L32},
  year={2023},
  publisher={IOP Publishing}
}

@article{bhatnagar2025fast,
  title={A Fast and Physically Grounded Ocean Model for GCMs: The Dynamical Slab Ocean Model of the Generic-PCM (rev. 3423)},
  author={Bhatnagar, Siddharth and Codron, Francis and Millour, Ehouarn and Bolmont, Emeline and Brunetti, Maura and Kasparian, J{\'e}r{\^o}me and Turbet, Martin and Chaverot, Guillaume},
  journal={EGUsphere},
  volume={2025},
  pages={1--41},
  year={2025},
  publisher={Copernicus Publications G{\"o}ttingen, Germany}
}

@article{cooke2023degenerate,
  title={Degenerate interpretations of O3 spectral features in exoplanet atmosphere observations due to stellar UV uncertainties: A 3D case study with TRAPPIST-1 e},
  author={Cooke, GJ and Marsh, DR and Walsh, Catherine and Youngblood, Allison},
  journal={The Astrophysical Journal},
  volume={959},
  number={1},
  pages={45},
  year={2023},
  publisher={IOP Publishing}
}

@article{carone2018stratosphere,
  title={Stratosphere circulation on tidally locked ExoEarths},
  author={Carone, Ludmila and Keppens, Rony and Decin, Leen and Henning, Th},
  journal={Monthly Notices of the Royal Astronomical Society},
  volume={473},
  number={4},
  pages={4672--4685},
  year={2018},
  publisher={Oxford University Press}
}

@ARTICLE{2025AJ....170..227H,
       author = {{Huber}, Philipp A. and {Dannert}, Felix A. and {Laugier}, Romain and {Matsuo}, Taro and {Rutten}, Loes W. and {Glauser}, Adrian M. and {Quanz}, Sascha P. and {LIFE Collaboration}},
        title = "{Robust Data Interpretation for Perturbed Nulling Interferometers via Proper Handling of Correlated Errors}",
      journal = {\aj},
         year = 2025,
        month = oct,
       volume = {170},
       number = {4},
          eid = {227},
        pages = {227},
          doi = {10.3847/1538-3881/adfb6b},
archivePrefix = {arXiv},
       eprint = {2508.15756},
 primaryClass = {astro-ph.IM},
       adsurl = {https://ui.adsabs.harvard.edu/abs/2025AJ....170..227H}
}

@ARTICLE{2025AJ....170..193D,
       author = {{Dannert}, Felix A. and {Huber}, Philipp A. and {Birbacher}, Thomas and {Laugier}, Romain and {Bonse}, Markus J. and {Garvin}, Emily O. and {Glauser}, Adrian M. and {Oehl}, Veronika and {Quanz}, Sascha P.},
        title = "{Consequences of Non-Gaussian Instrumental Noise in Perturbed Nulling Interferometers}",
      journal = {\aj},
         year = 2025,
        month = sep,
       volume = {170},
       number = {3},
          eid = {193},
        pages = {193},
          doi = {10.3847/1538-3881/add720},
archivePrefix = {arXiv},
       eprint = {2506.20653},
 primaryClass = {astro-ph.IM},
       adsurl = {https://ui.adsabs.harvard.edu/abs/2025AJ....170..193D}
}

@article{lewis2018influence,
  title={The influence of a substellar continent on the climate of a tidally locked exoplanet},
  author={Lewis, Neil T and Lambert, F Hugo and Boutle, Ian A and Mayne, Nathan J and Manners, James and Acreman, David M},
  journal={The Astrophysical Journal},
  volume={854},
  number={2},
  pages={171},
  year={2018},
  publisher={IOP Publishing}
}

@article{mak20243d,
  title={3D simulations of TRAPPIST-1e with varying CO2, CH4, and haze profiles},
  author={Mak, Mei Ting and Sergeev, Denis E and Mayne, Nathan and Banks, Nahum and Eager-Nash, Jake and Manners, James and Arney, Giada and H{\'e}brard, {\'E}ric and Kohary, Krisztian},
  journal={Monthly Notices of the Royal Astronomical Society},
  volume={529},
  number={4},
  pages={3971--3987},
  year={2024},
  publisher={Oxford University Press}
}

@article{mettler_earth_2024,
	title        = {Earth as an {Exoplanet}. {III}. {Using} {Empirical} {Thermal} {Emission} {Spectra} as an {Input} for {Atmospheric} {Retrieval} of an {Earth}-twin {Exoplanet}},
	author       = {Mettler, Jean-Noël and Konrad, Björn S. and Quanz, Sascha P. and Helled, Ravit},
	year         = 2024,
	month        = feb,
	journal      = {The Astrophysical Journal},
	volume       = 963,
	number       = 1,
	pages        = 24,
	doi          = {10.3847/1538-4357/ad198b},
	issn         = {0004-637X},
	url          = {https://dx.doi.org/10.3847/1538-4357/ad198b},
	urldate      = {2024-02-29},
	note         = {Publisher: The American Astronomical Society},
	language     = {en}
}

@article{alei_large_2022,
	title        = {Large {Interferometer} {For} {Exoplanets} ({LIFE}) - {V}. {Diagnostic} potential of a mid-infrared space interferometer for studying {Earth} analogs},
	author       = {Alei, Eleonora and Konrad, Björn S. and Angerhausen, Daniel and Grenfell, John Lee and Mollière, Paul and Quanz, Sascha P. and Rugheimer, Sarah and Wunderlich, Fabian},
	year         = 2022,
	month        = sep,
	journal      = {Astronomy \& Astrophysics},
	volume       = 665,
	pages        = {A106},
	doi          = {10.1051/0004-6361/202243760},
	issn         = {0004-6361, 1432-0746},
	url          = {https://www.aanda.org/articles/aa/abs/2022/09/aa43760-22/aa43760-22.html},
	urldate      = {2024-02-29},
	copyright    = {© E. Alei et al. 2022},
	note         = {Publisher: EDP Sciences},
	language     = {en}
}

@article{lobo2023terminator,
	title        = {Terminator habitability: the case for limited water availability on M-dwarf planets},
	author       = {Lobo, Ana H and Shields, Aomawa L and Palubski, Igor Z and Wolf, Eric},
	year         = 2023,
	journal      = {The Astrophysical Journal},
	publisher    = {IOP Publishing},
	volume       = 945,
	number       = 2,
	pages        = 161
}

@article{pearce2017origin,
	title        = {Origin of the RNA world: The fate of nucleobases in warm little ponds},
	author       = {Pearce, Ben KD and Pudritz, Ralph E and Semenov, Dmitry A and Henning, Thomas K},
	year         = 2017,
	journal      = {Proceedings of the National Academy of Sciences},
	publisher    = {National Academy of Sciences},
	volume       = 114,
	number       = 43,
	pages        = {11327--11332}
}

@article{sergeev_bistability_2022,
	title        = {Bistability of the {Atmospheric} {Circulation} on {TRAPPIST}-1e},
	author       = {Sergeev, Denis E. and Lewis, Neil T. and Lambert, F. Hugo and Mayne, Nathan J. and Boutle, Ian A. and Manners, James and Kohary, Krisztian},
	year         = 2022,
	month        = sep,
	journal      = {The Planetary Science Journal},
	volume       = 3,
	number       = 9,
	pages        = 214,
	doi          = {10.3847/PSJ/ac83be},
	issn         = {2632-3338},
	url          = {https://iopscience.iop.org/article/10.3847/PSJ/ac83be/meta},
	urldate      = {2023-05-09},
	note         = {Publisher: IOP Publishing},
	language     = {en}
}

@article{haqq-misra_demarcating_2018,
	title        = {Demarcating circulation regimes of synchronously rotating terrestrial planets within the habitable zone},
	author       = {Haqq-Misra, Jacob and Wolf, Eric T. and Joshi, Manoj and Zhang, Xi and Kopparapu, Ravi Kumar},
	year         = 2018,
	month        = jan,
	journal      = {The Astrophysical Journal},
	volume       = 852,
	number       = 2,
	pages        = 67,
	doi          = {10.3847/1538-4357/aa9f1f},
	issn         = {1538-4357},
	url          = {http://arxiv.org/abs/1710.00435},
	urldate      = {2021-02-24},
	note         = {arXiv: 1710.00435}
}

@article{carone_connecting_2015,
	title        = {Connecting the dots – {II}. {Phase} changes in the climate dynamics of tidally locked terrestrial exoplanets},
	author       = {Carone, L. and Keppens, R. and Decin, L.},
	year         = 2015,
	month        = nov,
	journal      = {Monthly Notices of the Royal Astronomical Society},
	volume       = 453,
	number       = 3,
	pages        = {2412--2437},
	doi          = {10.1093/mnras/stv1752},
	issn         = {0035-8711},
	url          = {https://doi.org/10.1093/mnras/stv1752},
	urldate      = {2022-03-17}
}

@article{chen_habitability_2019,
	title        = {Habitability and {Spectroscopic} {Observability} of {Warm} {M}-dwarf {Exoplanets} {Evaluated} with a {3D} {Chemistry}-{Climate} {Model}},
	author       = {Chen, Howard and Wolf, Eric T. and Zhan, Zhuchang and Horton, Daniel E.},
	year         = 2019,
	month        = nov,
	journal      = {The Astrophysical Journal},
	volume       = 886,
	number       = 1,
	pages        = 16,
	doi          = {10.3847/1538-4357/ab4f7e},
	issn         = {0004-637X},
	url          = {https://dx.doi.org/10.3847/1538-4357/ab4f7e},
	urldate      = {2022-12-03},
	note         = {Publisher: The American Astronomical Society},
	language     = {en}
}

@article{sergeev_trappist-1_2022,
	title        = {The {TRAPPIST}-1 {Habitable} {Atmosphere} {Intercomparison} ({THAI}). {II}. {Moist} {Cases}—{The} {Two} {Waterworlds}},
	author       = {Sergeev, Denis E. and Fauchez, Thomas J. and Turbet, Martin and Boutle, Ian A. and Tsigaridis, Kostas and Way, Michael J. and Wolf, Eric T. and Domagal-Goldman, Shawn D. and Forget, François and Haqq-Misra, Jacob and Kopparapu, Ravi K. and Lambert, F. Hugo and Manners, James and Mayne, Nathan J.},
	year         = 2022,
	month        = sep,
	journal      = {The Planetary Science Journal},
	volume       = 3,
	number       = 9,
	pages        = 212,
	doi          = {10.3847/PSJ/ac6cf2},
	issn         = {2632-3338},
	url          = {https://iopscience.iop.org/article/10.3847/PSJ/ac6cf2/meta},
	urldate      = {2022-09-22},
	note         = {Publisher: IOP Publishing},
	language     = {en}
}

@article{rutten_2025,
	journal = {Submitted to Astronomy \& Astrophysics},
	author = {Rutten, Loes W. and Laugier, Romain and Loicq Jérôme and Sigusch, Ida and Huber, Philipp A. and Dandumont, Colin and Dannert, Felix A. and Defrère, Denis},
	year   = 2025,
}

@article{fujii_probing_2025,
	title = {Probing thermal gradient of habitable-zone rocky planets with direct imaging as an anti-indicator of global surface ocean},
	journal = {Submitted to ApJ},
	author = {Fujii, Yuka and Angerhausen, Daniel and Matsuo, Taro and Wolf, Eric T.},
	year   = 2025,
}

@article{angerhausen_large_2024,
	title        = {Large {Interferometer} {For} {Exoplanets} ({LIFE}). {XII}. {The} {Detectability} of {Capstone} {Biosignatures} in the {Mid}-infrared—{Sniffing} {Exoplanetary} {Laughing} {Gas} and {Methylated} {Halogens}},
	author       = {Angerhausen, Daniel and Pidhorodetska, Daria and Leung, Michaela and Hansen, Janina and Alei, Eleonora and Dannert, Felix and Kammerer, Jens and Quanz, Sascha P. and Schwieterman, Edward W. and initiative, The LIFE},
	year         = 2024,
	month        = feb,
	journal      = {The Astronomical Journal},
	volume       = 167,
	number       = 3,
	pages        = 128,
	doi          = {10.3847/1538-3881/ad1f4b},
	issn         = {1538-3881},
	url          = {https://dx.doi.org/10.3847/1538-3881/ad1f4b},
	urldate      = {2024-05-13},
	note         = {Publisher: The American Astronomical Society},
	language     = {en}
}

@article{LIFE10,
	title        = {Large {Interferometer} {For} {Exoplanets} ({LIFE}) - {X}. {Detectability} of currently known exoplanets and synergies with future {IR}/{O}/{UV} reflected-starlight imaging missions},
	author       = {{Carri{\'o}n-Gonz{\'a}lez}, {\'O} and Kammerer, Jens and Angerhausen, Daniel and Dannert, Felix and Muñoz, Antonio García and Quanz, Sascha P. and Absil, Olivier and Beichman, Charles A. and Girard, Julien H. and Mennesson, Bertrand and Meyer, Michael R. and Stapelfeldt, Karl R.},
	year         = 2023,
	month        = oct,
	journal      = {Astronomy \& Astrophysics},
	volume       = 678,
	pages        = {A96},
	doi          = {10.1051/0004-6361/202347027},
	issn         = {0004-6361, 1432-0746},
	url          = {https://www.aanda.org/articles/aa/abs/2023/10/aa47027-23/aa47027-23.html},
	urldate      = {2024-05-13},
	copyright    = {© The Authors 2023},
	note         = {Publisher: EDP Sciences},
	language     = {en}
}

@article{LIFE9,
	title        = {{Large Interferometer For Exoplanets (LIFE). IX. Assessing the impact of clouds on atmospheric retrievals at mid-infrared wavelengths with a Venus-twin exoplanet}},
	author       = {{Konrad}, B.~S. and {Alei}, E. and {Quanz}, S.~P. and {Molli{\`e}re}, P. and {Angerhausen}, D. and {Fortney}, J.~J. and {Hakim}, K. and {Jordan}, S. and {Kitzmann}, D. and {Rugheimer}, S. and {Shorttle}, O. and {Wordsworth}, R. and {LIFE Collaboration}},
	year         = 2023,
	month        = may,
	journal      = {\aap},
	volume       = 673,
	pages        = {A94},
	doi          = {10.1051/0004-6361/202245655},
	eid          = {A94},
	archiveprefix = {arXiv},
	eprint       = {2303.04727},
	primaryclass = {astro-ph.EP},
	adsurl       = {https://ui.adsabs.harvard.edu/abs/2023A&A...673A..94K}
}

@article{LIFE8,
	title        = {{Large Interferometer for Exoplanets: VIII. Where Is the Phosphine? Observing Exoplanetary PH$_{3}$ with a Space-Based Mid-Infrared Nulling Interferometer}},
	author       = {{Angerhausen}, Daniel and {Ottiger}, Maurice and {Dannert}, Felix and {Miguel}, Yamila and {Sousa-Silva}, Clara and {Kammerer}, Jens and {Menti}, Franziska and {Alei}, Eleonora and {Konrad}, Bj{\"o}rn S. and {Wang}, Haiyang S. and {Quanz}, Sascha P. and {LIFE Collaboration}},
	year         = 2023,
	month        = feb,
	journal      = {Astrobiology},
	volume       = 23,
	number       = 2,
	pages        = {183--194},
	doi          = {10.1089/ast.2022.0010},
	archiveprefix = {arXiv},
	eprint       = {2211.04975},
	primaryclass = {astro-ph.EP},
	adsurl       = {https://ui.adsabs.harvard.edu/abs/2023AsBio..23..183A}
}

@article{LIFE6,
	title        = {{Large Interferometer For Exoplanets (LIFE). VI. Detecting rocky exoplanets in the habitable zones of Sun-like stars}},
	author       = {{Kammerer}, Jens and {Quanz}, Sascha P. and {Dannert}, Felix and {LIFE Collaboration}},
	year         = 2022,
	month        = dec,
	journal      = {\aap},
	volume       = 668,
	pages        = {A52},
	doi          = {10.1051/0004-6361/202243846},
	eid          = {A52},
	archiveprefix = {arXiv},
	eprint       = {2210.01782},
	primaryclass = {astro-ph.EP},
	adsurl       = {https://ui.adsabs.harvard.edu/abs/2022A&A...668A..52K}
}

@article{LIFE5,
	title        = {{Large Interferometer For Exoplanets (LIFE). V. Diagnostic potential of a mid-infrared space interferometer for studying Earth analogs}},
	author       = {{Alei}, Eleonora and {Konrad}, Bj{\"o}rn S. and {Angerhausen}, Daniel and {Grenfell}, John Lee and {Molli{\`e}re}, Paul and {Quanz}, Sascha P. and {Rugheimer}, Sarah and {Wunderlich}, Fabian and {LIFE Collaboration}},
	year         = 2022,
	month        = sep,
	journal      = {\aap},
	volume       = 665,
	pages        = {A106},
	doi          = {10.1051/0004-6361/202243760},
	eid          = {A106},
	archiveprefix = {arXiv},
	eprint       = {2204.10041},
	primaryclass = {astro-ph.EP},
	adsurl       = {https://ui.adsabs.harvard.edu/abs/2022A&A...665A.106A}
}

@article{LIFE3,
	title        = {{Large Interferometer For Exoplanets (LIFE). III. Spectral resolution, wavelength range, and sensitivity requirements based on atmospheric retrieval analyses of an exo-Earth}},
	author       = {{Konrad}, B.~S. and {Alei}, E. and {Quanz}, S.~P. and {Angerhausen}, D. and {Carri{\'o}n-Gonz{\'a}lez}, {\'O}. and {Fortney}, J.~J. and {Grenfell}, J.~L. and {Kitzmann}, D. and {Molli{\`e}re}, P. and {Rugheimer}, S. and {Wunderlich}, F. and {LIFE Collaboration}},
	year         = 2022,
	month        = aug,
	journal      = {\aap},
	volume       = 664,
	pages        = {A23},
	doi          = {10.1051/0004-6361/202141964},
	eid          = {A23},
	archiveprefix = {arXiv},
	eprint       = {2112.02054},
	primaryclass = {astro-ph.EP},
	adsurl       = {https://ui.adsabs.harvard.edu/abs/2022A&A...664A..23K}
}

@article{LIFE2,
	title        = {{Large Interferometer For Exoplanets (LIFE). II. Signal simulation, signal extraction, and fundamental exoplanet parameters from single-epoch observations}},
	author       = {{Dannert}, Felix A. and {Ottiger}, Maurice and {Quanz}, Sascha P. and {Laugier}, Romain and {Fontanet}, Emile and {Gheorghe}, Adrian and {Absil}, Olivier and {Dandumont}, Colin and {Defr{\`e}re}, Denis and {Gasc{\'o}n}, Carlos and {Glauser}, Adrian M. and {Kammerer}, Jens and {Lichtenberg}, Tim and {Linz}, Hendrik and {Loicq}, Jer{\^o}me and {LIFE Collaboration}},
	year         = 2022,
	month        = aug,
	journal      = {\aap},
	volume       = 664,
	pages        = {A22},
	doi          = {10.1051/0004-6361/202141958},
	eid          = {A22},
	archiveprefix = {arXiv},
	eprint       = {2203.00471},
	primaryclass = {astro-ph.EP},
	adsurl       = {https://ui.adsabs.harvard.edu/abs/2022A&A...664A..22D}
}

@article{LIFE1,
	title        = {{Large Interferometer For Exoplanets (LIFE). I. Improved exoplanet detection yield estimates for a large mid-infrared space-interferometer mission}},
	author       = {{Quanz}, S.~P. and {Ottiger}, M. and {Fontanet}, E. and {Kammerer}, J. and {Menti}, F. and {Dannert}, F. and {Gheorghe}, A. and {Absil}, O. and {Airapetian}, V.~S. and {Alei}, E. and {Allart}, R. and {Angerhausen}, D. and {Blumenthal}, S. and {Buchhave}, L.~A. and {Cabrera}, J. and {Carri{\'o}n-Gonz{\'a}lez}, {\'O}. and {Chauvin}, G. and {Danchi}, W.~C. and {Dandumont}, C. and {Defr{\'e}re}, D. and {Dorn}, C. and {Ehrenreich}, D. and {Ertel}, S. and {Fridlund}, M. and {Garc{\'\i}a Mu{\~n}oz}, A. and {Gasc{\'o}n}, C. and {Girard}, J.~H. and {Glauser}, A. and {Grenfell}, J.~L. and {Guidi}, G. and {Hagelberg}, J. and {Helled}, R. and {Ireland}, M.~J. and {Janson}, M. and {Kopparapu}, R.~K. and {Korth}, J. and {Kozakis}, T. and {Kraus}, S. and {L{\'e}ger}, A. and {Leedj{\"a}rv}, L. and {Lichtenberg}, T. and {Lillo-Box}, J. and {Linz}, H. and {Liseau}, R. and {Loicq}, J. and {Mahendra}, V. and {Malbet}, F. and {Mathew}, J. and {Mennesson}, B. and {Meyer}, M.~R. and {Mishra}, L. and {Molaverdikhani}, K. and {Noack}, L. and {Oza}, A.~V. and {Pall{\'e}}, E. and {Parviainen}, H. and {Quirrenbach}, A. and {Rauer}, H. and {Ribas}, I. and {Rice}, M. and {Romagnolo}, A. and {Rugheimer}, S. and {Schwieterman}, E.~W. and {Serabyn}, E. and {Sharma}, S. and {Stassun}, K.~G. and {Szul{\'a}gyi}, J. and {Wang}, H.~S. and {Wunderlich}, F. and {Wyatt}, M.~C. and {LIFE Collaboration}},
	year         = 2022,
	month        = aug,
	journal      = {\aap},
	volume       = 664,
	pages        = {A21},
	doi          = {10.1051/0004-6361/202140366},
	eid          = {A21},
	archiveprefix = {arXiv},
	eprint       = {2101.07500},
	primaryclass = {astro-ph.EP},
	adsurl       = {https://ui.adsabs.harvard.edu/abs/2022A&A...664A..21Q}
}

@article{LIFE13,
	title        = {Large Interferometer For Exoplanets (LIFE)-XIII. The value of combining thermal emission and reflected light for the characterization of Earth twins},
	author       = {Alei, Eleonora and Quanz, Sascha Patrick and Konrad, BS and Garvin, Emily Omaya and Kofman, Vincent and Mandell, Avi and Angerhausen, Daniel and Molli{\`e}re, Paul and Meyer, MR and Robinson, Tyler and others},
	year         = 2024,
	journal      = {Astronomy \& Astrophysics},
	publisher    = {EDP Sciences},
	volume       = 689,
	pages        = {A245}
}

@article{colose_effects_2021,
	title        = {Effects of {Spin}–{Orbit} {Resonances} and {Tidal} {Heating} on the {Inner} {Edge} of the {Habitable} {Zone}},
	author       = {Colose, Christopher M. and Haqq-Misra, Jacob and Wolf, Eric T. and Genio, Anthony D. Del and Barnes, Rory and Way, Michael J. and Ruedy, Reto},
	year         = 2021,
	month        = oct,
	journal      = {The Astrophysical Journal},
	volume       = 921,
	number       = 1,
	pages        = 25,
	doi          = {10.3847/1538-4357/ac135c},
	issn         = {0004-637X},
	url          = {https://dx.doi.org/10.3847/1538-4357/ac135c},
	urldate      = {2024-02-02},
	note         = {Publisher: The American Astronomical Society},
	language     = {en}
}

@article{shields_habitability_2016,
	title        = {The habitability of planets orbiting {M}-dwarf stars},
	author       = {Shields, Aomawa L. and Ballard, Sarah and Johnson, John Asher},
	year         = 2016,
	month        = dec,
	journal      = {Physics Reports},
	series       = {The habitability of planets orbiting {M}-dwarf stars},
	volume       = 663,
	pages        = {1--38},
	doi          = {10.1016/j.physrep.2016.10.003},
	issn         = {0370-1573},
	url          = {https://www.sciencedirect.com/science/article/pii/S0370157316303179},
	urldate      = {2022-02-27},
	language     = {en}
}

@article{LIFE12,
	title        = {{Large Interferometer For Exoplanets (LIFE). XII. The Detectability of Capstone Biosignatures in the Mid-infrared{\textemdash}Sniffing Exoplanetary Laughing Gas and Methylated Halogens}},
	author       = {{Angerhausen}, Daniel and {Pidhorodetska}, Daria and {Leung}, Michaela and {Hansen}, Janina and {Alei}, Eleonora and {Dannert}, Felix and {Kammerer}, Jens and {Quanz}, Sascha P. and {Schwieterman}, Edward W. and {The LIFE initiative}},
	year         = 2024,
	month        = mar,
	journal      = {\aj},
	volume       = 167,
	number       = 3,
	pages        = 128,
	doi          = {10.3847/1538-3881/ad1f4b},
	eid          = 128,
	adsurl       = {https://ui.adsabs.harvard.edu/abs/2024AJ....167..128A}
}

@article{tinetti_detectability_2006,
	title = {Detectability of {Planetary} {Characteristics} in {Disk}-{Averaged} {Spectra}. {I}: {The} {Earth} {Model}},
	volume = {6},
	issn = {1531-1074},
	shorttitle = {Detectability of {Planetary} {Characteristics} in {Disk}-{Averaged} {Spectra}. {I}},
	url = {https://ui.adsabs.harvard.edu/abs/2006AsBio...6...34T},
	doi = {10.1089/ast.2006.6.34},
	urldate = {2025-09-11},
	journal = {Astrobiology},
	author = {Tinetti, Giovanna and Meadows, Victoria S. and Crisp, David and Fong, William and Fishbein, Evan and Turnbull, Margaret and Bibring, Jean-Pierre},
	month = mar,
	year = {2006},
	pages = {34--47},
}

@article{tinetti_detectability_2006-1,
	title = {Detectability of {Planetary} {Characteristics} in {Disk}-{Averaged} {Spectra} {II}: {Synthetic} {Spectra} and {Light}-{Curves} of {Earth}},
	volume = {6},
	issn = {1531-1074},
	shorttitle = {Detectability of {Planetary} {Characteristics} in {Disk}-{Averaged} {Spectra} {II}},
	url = {https://ui.adsabs.harvard.edu/abs/2006AsBio...6..881T},
	doi = {10.1089/ast.2006.6.881},
	urldate = {2025-09-11},
	journal = {Astrobiology},
	author = {Tinetti, Giovanna and Meadows, Victoria S. and Crisp, David and Kiang, Nancy Y. and Kahn, Brian H. and Fishbein, Evan and Velusamy, Thangasamy and Turnbull, Margaret},
	month = dec,
	year = {2006},
	note = {ADS Bibcode: 2006AsBio...6..881T},
	pages = {881--900},
}

@article{kitzmann_clouds_2011,
	title = {Clouds in the atmospheres of extrasolar planets. {II}. {Thermal} emission spectra of {Earth}-like planets influenced by low and high-level clouds},
	volume = {531},
	issn = {0004-6361},
	url = {https://ui.adsabs.harvard.edu/abs/2011A&A...531A..62K},
	doi = {10.1051/0004-6361/201014343},
	urldate = {2025-09-11},
	journal = {Astronomy and Astrophysics},
	author = {Kitzmann, D. and Patzer, A. B. C. and von Paris, P. and Godolt, M. and Rauer, H.},
	month = jul,
	year = {2011},
	note = {ADS Bibcode: 2011A\&A...531A..62K},
	pages = {A62},
}

@article{des_marais_remote_2002,
	title = {Remote {Sensing} of {Planetary} {Properties} and {Biosignatures} on {Extrasolar} {Terrestrial} {Planets}},
	volume = {2},
	issn = {1531-1074},
	url = {https://ui.adsabs.harvard.edu/abs/2002AsBio...2..153D},
	doi = {10.1089/15311070260192246},
	urldate = {2025-09-11},
	journal = {Astrobiology},
	author = {Des Marais, David J. and Harwit, Martin O. and Jucks, Kenneth W. and Kasting, James F. and Lin, Douglas N. C. and Lunine, Jonathan I. and Schneider, Jean and Seager, Sara and Traub, Wesley A. and Woolf, Neville J.},
	month = jun,
	year = {2002},
	note = {ADS Bibcode: 2002AsBio...2..153D},
	pages = {153--181},
}

@article{hearty_mid-infrared_2009,
	title = {Mid-{Infrared} {Properties} of {Disk} {Averaged} {Observations} of {Earth} with {AIRS}},
	volume = {693},
	issn = {0004-637X},
	url = {https://ui.adsabs.harvard.edu/abs/2009ApJ...693.1763H},
	doi = {10.1088/0004-637X/693/2/1763},
	urldate = {2025-09-11},
	journal = {The Astrophysical Journal},
	author = {Hearty, Thomas and Song, Inseok and Kim, Sam and Tinetti, Giovanna},
	month = mar,
	year = {2009},
	note = {ADS Bibcode: 2009ApJ...693.1763H},
	pages = {1763--1774},
}

@article{gomez-leal_photometric_2012,
	title = {Photometric {Variability} of the {Disk}-integrated {Thermal} {Emission} of the {Earth}},
	volume = {752},
	issn = {0004-637X},
	url = {https://ui.adsabs.harvard.edu/abs/2012ApJ...752...28G},
	doi = {10.1088/0004-637X/752/1/28},
	urldate = {2025-09-11},
	journal = {The Astrophysical Journal},
	author = {Gómez-Leal, I. and Pallé, E. and Selsis, F.},
	month = jun,
	year = {2012},
	note = {ADS Bibcode: 2012ApJ...752...28G},
	pages = {28},
}

@article{olson2018atmospheric,
  title={Atmospheric seasonality as an exoplanet biosignature},
  author={Olson, Stephanie L and Schwieterman, Edward W and Reinhard, Christopher T and Ridgwell, Andy and Kane, Stephen R and Meadows, Victoria S and Lyons, Timothy W},
  journal={The Astrophysical Journal Letters},
  volume={858},
  number={2},
  pages={L14},
  year={2018},
  publisher={IOP Publishing}
}

@ARTICLE{2025A&A...701A.254K,
       author = {{Kozakis}, Thea and {Mendon{\c{c}}a}, Jo{\~a}o M. and {Buchhave}, Lars A. and {Lara}, Luisa M.},
        title = "{Is ozone a reliable proxy for molecular oxygen?: III. The impact of CH$_{4}$ on the O$_{2}${\textendash}O$_{3}$ relationship for Earth-like atmospheres}",
      journal = {\aap},
         year = 2025,
        month = sep,
       volume = {701},
          eid = {A254},
        pages = {A254},
          doi = {10.1051/0004-6361/202556015},
archivePrefix = {arXiv},
       eprint = {2508.19062},
 primaryClass = {astro-ph.EP},
       adsurl = {https://ui.adsabs.harvard.edu/abs/2025A&A...701A.254K}
}

@ARTICLE{2025A&A...699A.247K,
       author = {{Kozakis}, Thea and {Mendon{\c{c}}a}, Jo{\~a}o M. and {Buchhave}, Lars A. and {Lara}, Luisa M.},
        title = "{Is ozone a reliable proxy for molecular oxygen?: II. The impact of N$_{2}$O on the O$_{2}$-O$_{3}$ relationship for Earth-like atmospheres}",
      journal = {\aap},
         year = 2025,
        month = jul,
       volume = {699},
          eid = {A247},
        pages = {A247},
          doi = {10.1051/0004-6361/202555289},
archivePrefix = {arXiv},
       eprint = {2505.23279},
 primaryClass = {astro-ph.EP},
       adsurl = {https://ui.adsabs.harvard.edu/abs/2025A&A...699A.247K}
}

@inproceedings{glauser2024large,
  title={The Large Interferometer For Exoplanets (LIFE): a space mission for mid-infrared nulling interferometry},
  author={Glauser, Adrian M and Quanz, Sascha P and Hansen, Jonah and Dannert, Felix and Ireland, Michael and Linz, Hendrik and Absil, Olivier and Alei, Eleonora and Angerhausen, Daniel and Birbacher, Thomas and others},
  booktitle={Optical and Infrared Interferometry and Imaging IX},
  volume={13095},
  pages={354--374},
  year={2024},
  organization={SPIE}
}

@ARTICLE{2018AsBio..18..709C,
       author = {{Catling}, David C. and {Krissansen-Totton}, Joshua and {Kiang}, Nancy Y. and {Crisp}, David and {Robinson}, Tyler D. and {DasSarma}, Shiladitya and {Rushby}, Andrew J. and {Del Genio}, Anthony and {Bains}, William and {Domagal-Goldman}, Shawn},
        title = "{Exoplanet Biosignatures: A Framework for Their Assessment}",
      journal = {Astrobiology},
         year = 2018,
        month = jun,
       volume = {18},
       number = {6},
        pages = {709-738},
          doi = {10.1089/ast.2017.1737},
archivePrefix = {arXiv},
       eprint = {1705.06381},
 primaryClass = {astro-ph.EP},
       adsurl = {https://ui.adsabs.harvard.edu/abs/2018AsBio..18..709C}
}

@article{meadows2018exoplanet,
  title={Exoplanet biosignatures: understanding oxygen as a biosignature in the context of its environment},
  author={Meadows, Victoria S and Reinhard, Christopher T and Arney, Giada N and Parenteau, Mary N and Schwieterman, Edward W and Domagal-Goldman, Shawn D and Lincowski, Andrew P and Stapelfeldt, Karl R and Rauer, Heike and DasSarma, Shiladitya and others},
  journal={Astrobiology},
  volume={18},
  number={6},
  pages={630--662},
  year={2018},
  publisher={Mary Ann Liebert, Inc. 140 Huguenot Street, 3rd Floor New Rochelle, NY 10801 USA}
}

@ARTICLE{2017AJ....153...52K,
       author = {{Kane}, Stephen R. and {Gelino}, Dawn M. and {Turnbull}, Margaret C.},
        title = "{On the Orbital Inclination of Proxima Centauri b}",
      journal = {\aj},
         year = 2017,
        month = feb,
       volume = {153},
       number = {2},
          eid = {52},
        pages = {52},
          doi = {10.3847/1538-3881/153/2/52},
archivePrefix = {arXiv},
       eprint = {1612.02872},
 primaryClass = {astro-ph.EP},
       adsurl = {https://ui.adsabs.harvard.edu/abs/2017AJ....153...52K}
}

@article{fujii2018exoplanet,
  title={Exoplanet biosignatures: observational prospects},
  author={Fujii, Yuka and Angerhausen, Daniel and Deitrick, Russell and Domagal-Goldman, Shawn and Grenfell, John Lee and Hori, Yasunori and Kane, Stephen R and Pall{\'e}, Enric and Rauer, Heike and Siegler, Nicholas and others},
  journal={Astrobiology},
  volume={18},
  number={6},
  pages={739--778},
  year={2018},
  publisher={Mary Ann Liebert, Inc. 140 Huguenot Street, 3rd Floor New Rochelle, NY 10801 USA}
}

@article{paradise2022fundamental,
  title={Fundamental challenges to remote sensing of exo-earths},
  author={Paradise, Adiv and Menou, Kristen and Lee, Christopher and Fan, Bo Lin},
  journal={Monthly Notices of the Royal Astronomical Society},
  volume={512},
  number={3},
  pages={3616--3626},
  year={2022},
  publisher={Oxford University Press}
}

@article{konrad2024pursuing,
  title={Pursuing Truth: Improving Retrievals on Mid-infrared Exo-Earth Spectra with Physically Motivated Water Abundance Profiles and Cloud Models},
  author={Konrad, Bj{\"o}rn S and Quanz, Sascha P and Alei, Eleonora and Wordsworth, Robin},
  journal={The Astrophysical Journal},
  volume={975},
  number={1},
  pages={13},
  year={2024},
  publisher={IOP Publishing}
}

@article{barnes_tidal_2017,
    title = {Tidal locking of habitable exoplanets},
    volume = {129},
    issn = {0923-2958, 1572-9478},
    url = {http://link.springer.com/10.1007/s10569-017-9783-7},
    doi = {10.1007/s10569-017-9783-7},
    language = {en},
    number = {4},
    urldate = {2021-05-09},
    journal = {Celestial Mechanics and Dynamical Astronomy},
    author = {Barnes, Rory},
    month = dec,
    year = {2017},
    pages = {509--536},
}

@article{pierrehumbert_atmospheric_2019,
    title = {Atmospheric {Circulation} of {Tide}-{Locked} {Exoplanets}},
    volume = {51},
    url = {https://doi.org/10.1146/annurev-fluid-010518-040516},
    doi = {10.1146/annurev-fluid-010518-040516},
    number = {1},
    urldate = {2024-01-23},
    journal = {Annual Review of Fluid Mechanics},
    author = {Pierrehumbert, Raymond T. and Hammond, Mark},
    year = {2019},
    note = {\_eprint: https://doi.org/10.1146/annurev-fluid-010518-040516},
    pages = {275--303},
}

@article{renaud_tidal_2021,
    title = {Tidal {Dissipation} in {Dual}-body, {Highly} {Eccentric}, and {Nonsynchronously} {Rotating} {Systems}: {Applications} to {Pluto}–{Charon} and the {Exoplanet} {TRAPPIST}-1e},
    volume = {2},
    issn = {2632-3338},
    shorttitle = {Tidal {Dissipation} in {Dual}-body, {Highly} {Eccentric}, and {Nonsynchronously} {Rotating} {Systems}},
    url = {https://iopscience.iop.org/article/10.3847/PSJ/abc0f3/meta},
    doi = {10.3847/PSJ/abc0f3},
    language = {en},
    number = {1},
    urldate = {2024-08-26},
    journal = {The Planetary Science Journal},
    author = {Renaud, Joe P. and Henning, Wade G. and Saxena, Prabal and Neveu, Marc and Bagheri, Amirhossein and Mandell, Avi and Hurford, Terry},
    month = jan,
    year = {2021},
    note = {Publisher: IOP Publishing},
    pages = {4},
}

@article{sergeev_atmospheric_2020,
    title = {Atmospheric {Convection} {Plays} a {Key} {Role} in the {Climate} of {Tidally} {Locked} {Terrestrial} {Exoplanets}: {Insights} from {High}-resolution {Simulations}},
    volume = {894},
    issn = {0004-637X},
    shorttitle = {Atmospheric {Convection} {Plays} a {Key} {Role} in the {Climate} of {Tidally} {Locked} {Terrestrial} {Exoplanets}},
    url = {https://doi.org/10.3847/1538-4357/ab8882},
    doi = {10.3847/1538-4357/ab8882},
    language = {en},
    number = {2},
    urldate = {2021-09-17},
    journal = {The Astrophysical Journal},
    author = {Sergeev, Denis E. and Lambert, F. Hugo and Mayne, Nathan J. and Boutle, Ian A. and Manners, James and Kohary, Krisztian},
    month = may,
    year = {2020},
    note = {Publisher: American Astronomical Society},
    pages = {84},
}

@article{chen_biosignature_2018,
    title = {Biosignature {Anisotropy} {Modeled} on {Temperate} {Tidally} {Locked} {M}-dwarf {Planets}},
    volume = {868},
    issn = {2041-8205},
    url = {https://doi.org/10.3847/2041-8213/aaedb2},
    doi = {10.3847/2041-8213/aaedb2},
    language = {en},
    number = {1},
    urldate = {2020-12-20},
    journal = {The Astrophysical Journal},
    author = {Chen, Howard and Wolf, Eric T. and Kopparapu, Ravi and Domagal-Goldman, Shawn and Horton, Daniel E.},
    month = nov,
    year = {2018},
    note = {Publisher: American Astronomical Society},
    pages = {L6},
}

@article{cooke_variability_2023,
    title = {Variability due to climate and chemistry in observations of oxygenated {Earth}-analogue exoplanets},
    volume = {518},
    issn = {0035-8711},
    url = {https://doi.org/10.1093/mnras/stac2604},
    doi = {10.1093/mnras/stac2604},
    number = {1},
    urldate = {2024-03-07},
    journal = {Monthly Notices of the Royal Astronomical Society},
    author = {Cooke, G J and Marsh, D R and Walsh, C and Rugheimer, S and Villanueva, G L},
    month = jan,
    year = {2023},
    pages = {206--219},
}

@article{cohen2023traveling,
  title={Traveling planetary-scale waves cause cloud variability on tidally locked aquaplanets},
  author={Cohen, Maureen and Bollasina, Massimo A and Sergeev, Denis E and Palmer, Paul I and Mayne, Nathan J},
  journal={The Planetary Science Journal},
  volume={4},
  number={4},
  pages={68},
  year={2023},
  publisher={IOP Publishing}
}

@article{hochman2022greater,
  title={Greater climate sensitivity and variability on TRAPPIST-1e than Earth},
  author={Hochman, Assaf and De Luca, Paolo and Komacek, Thaddeus D},
  journal={The Astrophysical Journal},
  volume={938},
  number={2},
  pages={114},
  year={2022},
  publisher={IOP Publishing}
}

@article{chen_effects_2025,
  title={Effects of Transient Stellar Emissions on Planetary Climates of Tidally Locked Exo-Earths},
  author={Chen, Howard and De Luca, Paolo and Hochman, Assaf and Komacek, Thaddeus D},
  journal={The Astronomical Journal},
  volume={170},
  number={1},
  pages={40},
  year={2025},
  publisher={IOP Publishing}
}

@article{chen2021persistence,
  title={Persistence of flare-driven atmospheric chemistry on rocky habitable zone worlds},
  author={Chen, Howard and Zhan, Zhuchang and Youngblood, Allison and Wolf, Eric T and Feinstein, Adina D and Horton, Daniel E},
  journal={Nature Astronomy},
  volume={5},
  number={3},
  pages={298--310},
  year={2021},
  publisher={Nature Publishing Group UK London}
}

@article{way2017effects,
  title={Effects of variable eccentricity on the climate of an Earth-like world},
  author={Way, Michael J and Georgakarakos, Nikolaos},
  journal={The Astrophysical Journal Letters},
  volume={835},
  number={1},
  pages={L1},
  year={2017},
  publisher={IOP Publishing}
}

@article{kozakis_is_2022,
  title={Is ozone a reliable proxy for molecular oxygen?-I. The O2--O3 relationship for Earth-like atmospheres},
  author={Kozakis, Thea and Mendon{\c{c}}a, Jo{\~a}o M and Buchhave, Lars A},
  journal={Astronomy \& Astrophysics},
  volume={665},
  pages={A156},
  year={2022},
  publisher={EDP Sciences}
}

@article{schwieterman_exoplanet_2018,
    title = {Exoplanet {Biosignatures}: {A} {Review} of {Remotely} {Detectable} {Signs} of {Life}},
    volume = {18},
    issn = {1531-1074, 1557-8070},
    shorttitle = {Exoplanet {Biosignatures}},
    url = {http://www.liebertpub.com/doi/10.1089/ast.2017.1729},
    doi = {10.1089/ast.2017.1729},
    language = {en},
    number = {6},
    urldate = {2020-11-09},
    journal = {Astrobiology},
    author = {Schwieterman, Edward W. and Kiang, Nancy Y. and Parenteau, Mary N. and Harman, Chester E. and DasSarma, Shiladitya and Fisher, Theresa M. and Arney, Giada N. and Hartnett, Hilairy E. and Reinhard, Christopher T. and Olson, Stephanie L. and Meadows, Victoria S. and Cockell, Charles S. and Walker, Sara I. and Grenfell, John Lee and Hegde, Siddharth and Rugheimer, Sarah and Hu, Renyu and Lyons, Timothy W.},
    month = jun,
    year = {2018},
    pages = {663--708},
}

@article{noda_circulation_2017,
    title = {The circulation pattern and day-night heat transport in the atmosphere of a synchronously rotating aquaplanet: {Dependence} on planetary rotation rate},
    volume = {282},
    issn = {0019-1035},
    shorttitle = {The circulation pattern and day-night heat transport in the atmosphere of a synchronously rotating aquaplanet},
    url = {https://www.sciencedirect.com/science/article/pii/S0019103516305668},
    doi = {10.1016/j.icarus.2016.09.004},
    urldate = {2023-12-11},
    journal = {Icarus},
    author = {Noda, S. and Ishiwatari, M. and Nakajima, K. and Takahashi, Y. O. and Takehiro, S. and Onishi, M. and Hashimoto, G. L. and Kuramoto, K. and Hayashi, Y. -Y.},
    month = jan,
    year = {2017},
    pages = {1--18},
}

@article{bhongade_asymmetries_2024,
  title={Asymmetries in the Simulated Ozone Distribution on TRAPPIST-1e due to Orography},
  author={Bhongade, Anand and Marsh, Daniel R and Sainsbury-Martinez, Felix and Cooke, Gregory},
  journal={The Astrophysical Journal},
  volume={977},
  number={1},
  pages={96},
  year={2024},
  publisher={IOP Publishing}
}

@ARTICLE{2025ApJ...982L...2L,
       author = {{Leung}, Michaela and {Tsai}, Shang-Min and {Schwieterman}, Edward W. and {Angerhausen}, Daniel and {Hansen}, Janina},
        title = "{Examining the Potential for Methyl Halide Accumulation and Detectability in Possible Hycean-type Atmospheres}",
      journal = {\apjl},
         year = 2025,
        month = mar,
       volume = {982},
       number = {1},
          eid = {L2},
        pages = {L2},
          doi = {10.3847/2041-8213/adb558},
archivePrefix = {arXiv},
       eprint = {2502.13856},
 primaryClass = {astro-ph.EP},
       adsurl = {https://ui.adsabs.harvard.edu/abs/2025ApJ...982L...2L}
}
\begin{appendix}
\section{Hemispheric mean climate-chemistry model output}
In Tables~\ref{tab:mean_3d_diagnostics_11} and \ref{tab:mean_3d_diagnostics_32}, we present the hemispheric mean climate and chemistry diagnostics, as a function of phase angle and averaged over the observed hemisphere. These are visualised in Figure~\ref{fig:phase_ccm_diags}.
\begin{table}[!htbp]
\centering
\caption{Hemispheric means across the observed hemisphere of Proxima Centauri b in 1:1 SOR.}
\small
\begin{tabular}{c|c|c|c|c|c}
\hline
$\theta(t)$ & $\overline{T_{S}}$ & $\overline{CWP}$ & $\overline{\sigma_{H2O}}$ & $\overline{\sigma_{O3}}$ & $\overline{\chi_{O3, Strat}}$ \\
{\scriptsize ($^\circ$)} & {\scriptsize (K)} & {\scriptsize ($kg~m^{-2}$)} & {\scriptsize ($molec~m^{-2}$)} & {\scriptsize ($molec~m^{-2}$)} & {\scriptsize ($mol~mol^{-1}$)} \\
\hline
168.26 & 200.15 & 5.99e-4 & 1.46e+25 & 1.29e+23 & 8.95e-6 \\
200.45 & 201.46 & 5.58e-4 & 1.54e+25 & 1.30e+23 & 8.96e-6 \\
232.63 & 212.20 & 1.82e-2 & 3.79e+25 & 1.27e+23 & 8.98e-6 \\
264.81 & 226.07 & 5.52e-2 & 1.03e+26 & 1.17e+23 & 9.01e-6 \\
297.00 & 242.14 & 1.09e-1 & 1.87e+26 & 9.76e+22 & 9.03e-6 \\
329.18 & 255.16 & 1.29e-1 & 2.21e+26 & 8.41e+22 & 9.01e-6 \\
1.36 & 260.28 & 1.25e-1 & 2.27e+26 & 7.91e+22 & 8.98e-6 \\
33.55 & 254.72 & 1.21e-1 & 2.25e+26 & 7.88e+22 & 8.99e-6 \\
65.73 & 241.27 & 1.04e-1 & 1.91e+26 & 8.47e+22 & 8.96e-6 \\
97.91 & 224.82 & 5.35e-2 & 1.03e+26 & 1.02e+23 & 8.93e-6 \\
130.09 & 209.69 & 1.12e-2 & 3.42e+25 & 1.21e+23 & 8.95e-6 \\
162.28 & 200.27 & 8.93e-4 & 1.69e+25 & 1.29e+23 & 8.95e-6 \\
\hline
\end{tabular}
\label{tab:mean_3d_diagnostics_11}
\tablefoot{Contains the quantities shown in Figure~\ref{fig:pcb_3dchem_distrib_phase} and the mean volume mixing ratio of O$_3$ in the stratosphere.}
\end{table}
\begin{table}[!htbp]
\centering
\caption{Hemispheric means across the observed hemisphere of Proxima Centauri b in 3:2 SOR.}
\small
\begin{tabular}{c|c|c|c|c|c}
\hline
$\theta(t)$ & $\overline{T_{S}}$ & $\overline{CWP}$ & $\overline{\sigma_{H2O}}$ & $\overline{\sigma_{O3}}$ & $\overline{\chi_{O3, Strat}}$ \\
{\scriptsize ($^\circ$)} & {\scriptsize (K)} & {\scriptsize ($kg~m^{-2}$)} & {\scriptsize ($molec~m^{-2}$)} & {\scriptsize ($molec~m^{-2}$)} & {\scriptsize ($mol~mol^{-1}$)} \\
\hline
261.57 & 260.57 & 1.19e-1 & 1.38e+26 & 1.98e+23 & 1.15e-5 \\
280.15 & 261.74 & 1.08e-1 & 1.38e+26 & 1.92e+23 & 1.14e-5 \\
298.10 & 259.98 & 9.54e-2 & 1.23e+26 & 1.90e+23 & 1.13e-5 \\
318.57 & 257.26 & 6.78e-2 & 9.22e+25 & 1.92e+23 & 1.14e-5 \\
341.73 & 256.08 & 4.66e-2 & 7.85e+25 & 1.95e+23 & 1.14e-5 \\
11.98 & 257.60 & 4.76e-2 & 8.99e+25 & 1.96e+23 & 1.15e-5 \\
58.37 & 259.34 & 5.55e-2 & 1.13e+26 & 1.97e+23 & 1.15e-5 \\
118.08 & 259.03 & 7.18e-2 & 1.15e+26 & 2.00e+23 & 1.14e-5 \\
172.81 & 257.49 & 6.50e-2 & 9.29e+25 & 2.04e+23 & 1.13e-5 \\
210.38 & 256.46 & 6.40e-2 & 8.36e+25 & 2.06e+23 & 1.13e-5 \\
236.02 & 258.00 & 7.55e-2 & 9.89e+25 & 2.04e+23 & 1.13e-5 \\
257.80 & 260.53 & 1.16e-1 & 1.38e+26 & 1.98e+23 & 1.14e-5 \\
276.86 & 261.58 & 1.06e-1 & 1.41e+26 & 1.92e+23 & 1.14e-5 \\
294.62 & 260.17 & 9.98e-2 & 1.32e+26 & 1.89e+23 & 1.15e-5 \\
314.56 & 257.44 & 6.49e-2 & 1.02e+26 & 1.91e+23 & 1.15e-5 \\
337.14 & 255.90 & 4.75e-2 & 8.01e+25 & 1.94e+23 & 1.14e-5 \\
5.35 & 257.44 & 4.58e-2 & 8.82e+25 & 1.96e+23 & 1.13e-5 \\
48.36 & 259.53 & 5.13e-2 & 1.10e+26 & 1.97e+23 & 1.12e-5 \\
106.72 & 259.45 & 7.13e-2 & 1.19e+26 & 1.99e+23 & 1.14e-5 \\
163.82 & 257.84 & 6.33e-2 & 9.50e+25 & 2.03e+23 & 1.16e-5 \\
204.62 & 256.26 & 5.84e-2 & 8.49e+25 & 2.07e+23 & 1.16e-5 \\
231.68 & 257.16 & 6.35e-2 & 9.32e+25 & 2.06e+23 & 1.16e-5 \\
253.94 & 259.78 & 1.23e-1 & 1.35e+26 & 2.00e+23 & 1.15e-5 \\
273.52 & 261.51 & 1.06e-1 & 1.41e+26 & 1.93e+23 & 1.13e-5 \\
\hline
\end{tabular}
\label{tab:mean_3d_diagnostics_32}
\tablefoot{Contains the the quantities shown in Figure~\ref{fig:pcb_3dchem_distrib_phase} and the mean volume mixing ratio of O$_3$ in the stratosphere. The results are shown for two orbits.}
\end{table}
\newline

%\clearpage
\section{Generalisation to other targets}\label{app:brighter_stars}
\begin{figure}
    \centering
    
    \includegraphics[width=0.89\linewidth]{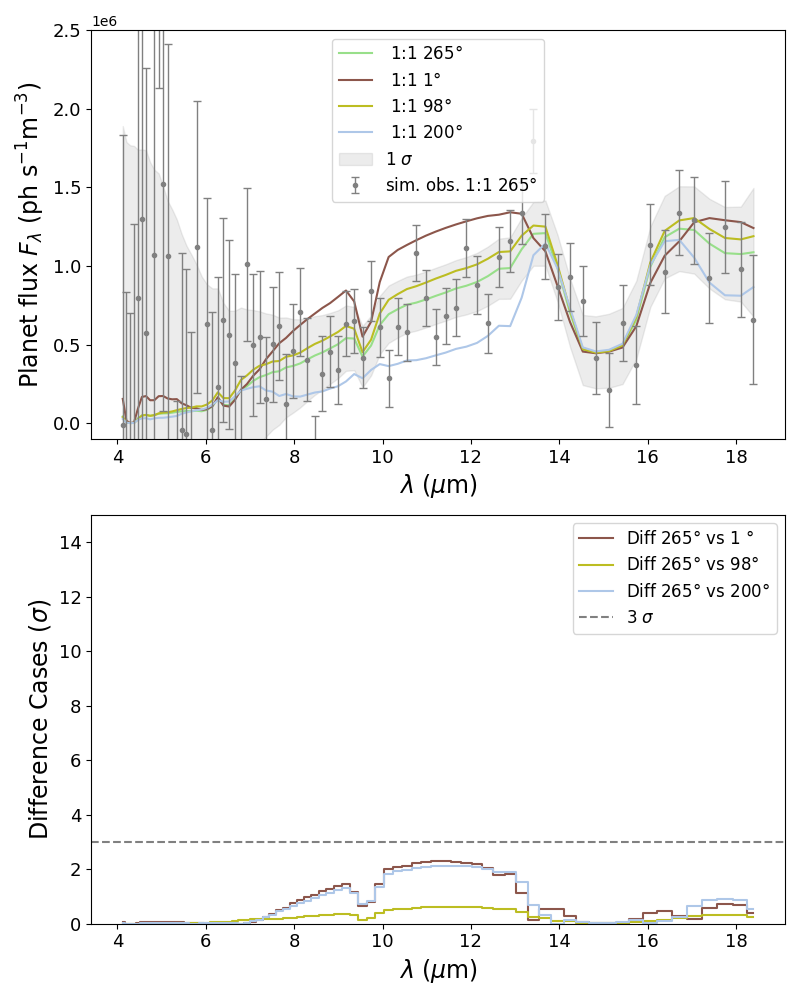}

    \caption{Same as Figure~\ref{fig:LIFE_spectra_all} but for a M3V system at 5 pcs and 4 days of observation time. Note the one order of magnitude smaller scale on the flux axis. The grey area represents the 1-$\sigma$ sensitivity; the dark grey error bars show an individual simulated observation. Lower panel: Statistical significance of the detected differences between different phases.}
    \label{fig:LIFE_spectra_all_M3V}
\end{figure}

As mentioned in Section~\ref{sec:disc_LIFE_observing}, we performed initial tests to investigate whether analogous variability is accessible around brighter
stars. We use the 4D distributions from the Proxima Centauri b configurations (Section~\ref{subsec:3d_results}) to create synthetic LIFE spectra for exoplanets in the HZ of brighter stars. Figure~\ref{fig:LIFE_spectra_all_M3V} shows the phase angle dependence of these spectra around an M3V star (3439 K, 0.361 $R_{Sun}$, 0.127 AU) at 5~pcs. Such targets represent a more typical early type LIFE target \citep{angerhausen_large_2024, 2025ApJ...982L...2L}, with dozens of observable planets at favourable inclination. From Kepler's third law, such an exoplanet covers an orbital phase angle change of ${\sim}$32$^\circ$ in four days, which we take as the observation time. Whilst the 4D variations are not as clearly distinguishable as in the golden target case of Proxima Centauri b, phase angle variations are of the same order of magnitude as the noise floor visible and may still provide a source of confusion when taking snapshot spectra. We note that these tests are not physically or chemically consistent, as the distinct orbital, dynamical, and photochemical timescales likely produce a different 4D atmospheric state. A more detailed analysis of these early type LIFE targets is presented in \citet{fujii_probing_2025}. We suggest that future work focus on self-consistently simulating the 4D atmospheric chemistry and exploring observational prospects for these targets. A crucial aspect will be the comparison of orbital and chemical timescales, to investigate whether longer observations still probe 4D chemistry or whether these smooth out any variations.

% -same models for HZ planet in M3V system at 5pcs (3439 K, 0.361 $R_{Sun}$, 0.127 AU, ... \textbf{[CHECKME]}

% -4 days of observation similar fraction of orbital period as in our Prox B case

% -caveat: just Prox B models placed there, different photechem/orb. timescales etc. not included

% -more "typical" early type LIFE target (Angerhausen+ 2024, Leung+ 2025) with dozens of expected planets observable, many of those with favorable inclinations

% -mode detailed analysis on this in (Fuji+ 2025, in prep)

% -in summary: while not as clearly distinguishable as in the "golden Target" case of Prox B it still seems to be measurable and/or similar source for possible confusion/general calibration issues

\end{appendix}

\end{document}